\documentclass[12pt]{article}
\usepackage{amsmath}
\usepackage{amssymb}
\begin{document}
\begin{center}
{\bf {\Large   {Superfield Approach to Nilpotency and Absolute Anticommutativity of Conserved Charges: 2D Non-Abelian 1-Form Gauge Theory}}}

\vskip 3.0cm

{\sf S. Kumar$^{(a)}$, B. Chauhan$^{(a)}$, R. P. Malik$^{(a,b)}$}\\
$^{(a)}$ {\it Physics Department, Center of Advance Studies, Institute of Science,}\\
{\it Banaras Hindu University, Varanasi - 221 005, (U.P.), India}\\

\vskip 0.1cm

\vskip 0.1cm

$^{(b)}$ {\it DST Centre for Interdisciplinary Mathematical Sciences,}\\
{\it Institute of Science, Banaras Hindu University, Varanasi - 221 005, India}\\
{\small {\sf {e-mails: sunil.bhu93@gmail.com; bchauhan501@gmail.com;  rpmalik1995@gmail.com}}}

\end{center}

\vskip 2cm

\noindent
{\bf Abstract:} We exploit the theoretical strength of augmented version of superfield approach (AVSA) to 
Becchi-Rouet-Stora-Tyutin (BRST) formalism to express the nilpotency and absolute anticommutativity 
properties of the (anti-)BRST and (anti-)co-BRST conserved charges
for the two $(1+1)$-dimensional (2D) non-Abelian 1-form gauge theory (without any interaction with matter fields)
in the language of superspace variables, their derivatives  and suitable superfields.
 In the proof of absolute anticommutativity property, we invoke the strength of Curci-Ferrari (CF)
condition  for the (anti-)BRST charges. No such {\it outside} condition/restriction
 is required in the proof of absolute anticommutativity 
of the (anti-)co-BRST conserved charges. 
The latter observation (as well as other observations) connected with
(anti-)co-BRST  symmetries and corresponding  conserved charges are  {\it novel} results of our present investigation.
We also discuss the (anti-)BRST and (anti-)co-BRST symmetry invariance of the appropriate Lagrangian densities 
 within the framework of AVSA. In addition, we dwell a bit on the derivation of the above fermionic (nilpotent) symmetries by
 applying the AVSA to BRST formaism where {\it only} the (anti-)chiral superfields are used. 

\vskip 2.0cm

\noindent
{\it {Keywords}}: {2D non-Abelian $1$-form gauge theory; augmented superfield formalism;
(anti-) chiral superfields; nilpotency and 
absolute anticommutativity properties; (anti-)BRST and (anti-)co-BRST   
conserved charges;  symmetry 
invariance; geometrical interpretation

\newpage
\section{Introduction}

The  principle of {\it local} gauge invariance is at the heart of
 standard model of particle physics where 
there is a stunning degree of agreement between theory and experiment.
 One of the most elegant approaches to covariantly
quantize the above kinds of gauge theories (based on the principle of {\it local}
 gauge invariance) is Becchi-Rouet-Stora-Tyutin (BRST)
formalism where {\it each} local gauge symmetry 
is traded with {\it two} nilpotent symmetries which are christened 
as the BRST and anti-BRST symmetries. The latter symmetries are the {\it quantum} version of 
the gauge symmetries and their very existence 
ensures the covariant canonical quantization of a given gauge theory. 
The decisive features of the above quantum (anti-)BRST symmetries 
are the observations that (i) they are nilpotent of order two, and (ii) 
they are absolutely anticommuting in nature. In the language
of theoretical physics, the nilpotency property ensures the fermionic 
(supersymmetic-type) nature of the (anti-)BRST symmetries and 
the linear independence of BRST and anti-BRST symmetries is 
encoded in  the property of absolute anticommutativity of the above (anti-)BRST symmetries.

The superfield approach to BRST formalism [1-8] provides 
the geometrical basis for the properties of 
nilpotency and absolute anticommutativity that are 
associated with the (anti-)BRST symmetries. In the above 
{\it usual}  superfield approach [1-8], the celebrated 
horizontality condition (HC) plays a key and 
decisive role. The HC leads, however, to the 
derivation (as well as geometrical interpretation) of the (anti-)BRST 
symmetries that are associated with the gauge and 
corresponding (anti-)ghost fields {\it only }. 
It does {\it not} shed any light on the (anti-)BRST
 symmetries that are associated with the {\it matter }fields 
in a given {\it interacting } gauge theory.
 In a set of papers [9-12], the above {\it usual } superfield 
formalism has been systematically generalized so
as to derive the (anti-)BRST symmetries for the gauge,
 {\it matter} and (anti-)ghost fields {\it together}.
The latter superfield approach [9-12] has been christened as the augmented 
version of superfield approach to BRST formalism where, consistent
with the HC, additional restrictions (i.e. gauge invariant conditions) 
are also invoked. We shall exploit the latter superfield approach [9-12] to 
discuss a few key features of the 2D non-Abelian 1-form 
gauge theory (without any interaction with matter fields) that have already been discussed 
within the framework of BRST formalism [13-16].

To be more specific, in the above works [13-16], we have shown the existence
of the nilpotent (anti-)BRST as well as (anti-)co-BRST symmetry  
transformations for the 2D non-Abelian 
1-form gauge theory.
The central theme of our present investigation is to capture the nilpotency 
and absolute anticommutativity of the (anti-)BRST and (anti-)co-BRST 
conserved charges [13-16] within the framework of augmented version of 
superfield approach (AVSA) to BRST formalism. In the proof of the absolute
 anticommutativity of the (anti-)BRST charges (within the framework of AVSA),
 we invoke the CF-condition to recast the expressions for these charges in an
 appropriate form and, then only, the superfield formalism is applied. However,
 in the case of the {\it above} proof of the (anti-)co-BRST charges, we do {\it not}
 invoke any CF-type restrictions. In our present investigation, we have proven 
the nilpotency and absolute anticommutativity of the conserved (anti-)BRST and 
(anti-) co-BRST charges that have been derived from {\it two} sets of coupled 
Lagrangian densities (cf. Eqs. (1), (39) below) for our 2D non-Abelian 1-form 
theory.

In the BRST approach to a given gauge theory, the existence of the (anti-)BRST
symmetries and their conserved charges is well-known. However, we have been
able to establish the existence of (anti-)co-BRST symmetry transformations
(in addition to the nilpotent (anti-)BRST symmetry transformations) in the case of a toy model
of a rigid rotor in one (0 + 1)-dimension of  spacetime [17].
Furthermore, we have demonstrated the existence of such (i.e. (anti-)co-BRST)
symmetries in the cases of Abelian $p$-form ($p$ = 1, 2, 3)  gauge theories 
in the two (1+1)-dimensions, four (3+1)-dimensions and six (5+1)-dimensions of spacetime
(see, e.g., [18] and reference therein). In other words, we have established
that the nilpotent (anti-)co-BRST symmetries  exist for any arbitrary Abelian $p$-form
($p$ = 1, 2, 3,...)  gauge theory  in $D = 2p$ dimensions of spacetime [18].
One of the decisive features of the (anti-)co-BRST symmetries is the observation that
it is the gauge-fixing term that remains invariant under these transformations
 (unlike the kinetic term 
that remains invariant under the (anti-)BRST  transformations).
The geometrical origin for   these observations has been provided in our 
review article (see, e.g. [18] and reference therein).

We concentrate on the 2D non-Abelian theory (without any interaction with matter fields)
 because this theory has been shown [13] to be a perfect 
model of Hodge theory as well as a {\it new} model of topological 
field theory (TFT) which captures a few aspects of Witten-type TFTs [19] and some
salient features of Schwarz-type TFTs [20].   
The equivalence of the coupled Lagrangian densities of 
this 2D theory with respect to the (anti-)co-BRST symmetries 
has been established in our recent 
publication [14]. We have also discussed the CF-type restrictions 
for this theory within the framework of superfield approach [15] where we have demonstrated 
the existence of a tower of CF-type restrictions. This happens for {\it this} theory 
because it is a TFT where there are {\it no} physical 
propagating  degrees of freedom 
for the 2D gauge field. In an another work [16], we have derived 
all the conserved currents and charges for this 2D theory and shown their algebraic 
structure that is found to be reminiscent of the Hodge algebra [21-24]. In other words,
we have provided the physical realizations of the de Rham cohomological operators
of differential geometry (and their algebra) in the language of 
the continuous (as well as discrete) symmetries, corresponding  conserved charges and their algebra
in operator form.

We have exploited the key ideas of AVSA to BRST formalism to derive 
the (anti-) BRST and (anti-)co-BRST symmetry transformations by using
HC and dual-HC (DHC) as well as the (anti-)chiral superfield approach
to BRST formalism (see, Appendices A and B below) in the context of our
present 2D non-Abelian 1-form gauge theory. In our earlier 
works [9-12], we have {\it never} been able to capture the nilpotency
as well as absolute anticommutativity properties of the 
(anti-)BRST and (anti-)co-BRST charges. The central objective 
of our present paper is to achieve this goal in the case of 2D
non-Abelian 1-form gauge theory.
To the best of our knowledge, this issue is being pursued  
for the first time in our present endeavor. Thus, the {\it novelty} in our present 
investigation is the observation that the nilpotency of the fermionic
symmetry transformations and CF-type restrictions 
play a decisive role  in capturing the 
nilpotency and absolute anticommutativity properties  of the
conserved (anti-)BRST and (anti-)co-BRST charges in the ordinary 2D spacetime (see, Sec. 5 below).
However, it is the nilpotency of the translational generators (along the Grassmannian directions) that plays
a crucial role for the {\it same} purpose within the framework of AVSA to BRST formalism on the supermanifold
(see, Sec. 6 below).

The following key factors have spurred our curiosity to pursue our present
investigation. First, to add some {\it new} ideas to the existing 
technique(s) of the superfield formalism is a challenging problem. In this context,
we have expressed the fermionic charges  (i.e. nilpotent (anti-)BRST and (anti-)co-BRST) charges
in the language of the superfields and derivatives defined on the (2, 2)-dimensional
supermanifold. Second, in our earlier works [14,15], we have derived the
expressions for the conserved fermionic charges  in the ordinary 2D space.
It is a challenging  problem to express their nilpotency and absolute 
anticommutativity properties in terms of the quantities that are defined on 
the (2, 2)-dimensional supermanifold. Third, it is also an interesting 
as well as {\it novel} idea
to discuss various aspects of the (anti-)co-BRST charges within the
framework of AVSA to BRST formalism.
Finally, the insights and 
understandings, gained in our present investigation, would turn out to 
be useful when we shall discuss the 4D Abelian 2-form and 6D Abelian 3-form 
gauge theories within the framework of AVSA to BRST
 formalism. In fact, we have already shown, in our earlier works 
[25,26], that the above 4D and 6D  Abelian
2-form and 3-form gauge theories are the models for the Hodge theory 
and they {\it do} support the existence of the (anti-)BRST and 
(anti-)co-BRST symmetries (as well as their corresponding conserved charges)
in addition to the {\it other} continuous symmetries (and corresponding charges).
There exist discrete symmetries, too, in these theories [25,26]. All these symmetries (and corresponding conserved charges)
are required for the proof that the above models are the tractable field theoretic examples of Hodge theory.

Our present paper is organized  as follows. In Sec. 2, we discuss the nilpotent
 (fermionic) (anti-)BRST and (anti-)co-BRST symmetries in the 
Lagrangian formulation. Our Sec. 3 is devoted to the discussion of horizontality 
condition (HC) that leads to the derivation of  (anti-)BRST symmetries for the
 gauge field and
corresponding fermionic (anti-)ghost fields along with the CF-condition. Sec. 4 of 
our paper deals with the dual-HC (DHC) which enables us to derive the (anti-)co-BRST 
symmetries that exist for the 2D non-Abelian 1-form gauge theory. 
The subject matter of Sec. 5 concerns 
itself with the discussion of nilpotency and absolute
anticommutativity properties of the fermionic charges within the 
framework of BRST formalism in 2D ordinary spacetime. In Sec. 6, we discuss the nilpotency and
absolute anticommutativity of the fermionic charges  
within the framework of  AVSA to BRST formalism on a (2, 2)-dimensional
supermanifold
 where the CF-condition plays an important role for (anti-)BRST charges. 
Finally, we discuss the key results of our present 
investigation in Sec. 7 where we point out a few possible theoretical directions
which might be pursued for  future investigations.

In our Appendices A and B, we derive the (anti-)BRST and (anti-)co-BRST symmetry transformations by exploiting
the ideas of (anti-)chiral superfield approach to BRST formalism
which match with the {\it ones} derived in the main body of the text. We express the
(anti-)BRST and (anti-)co-BRST invariance of the Lagrangian densities in the language of
the AVSA to BRST formalism in our Appendix C.

We note that the
theoretical materials, contained in Secs. 5 and 6, are deeply inter-related. In fact, sometimes, it is due to
our observations in Sec. 5 that we have been able to express the nilpotency
and anticommutativity properties of the charges in Sec. 6 within the framework of 
AVSA to BRST formalism. On the other hand, at times, it is our knowledge of the AVSA to BRST formalism (cf. Sec. 6)
that has turned out to be handy for our derivations of the above properties in 2D ordinary space (cf. Sec. 5).\\

\noindent
{\it Convention and Notations}: We take the 2D ordinary Minkowskian background spacetime to be {\it flat} 
with a metric tensor $\eta_{\mu\nu} = \;$diag$\;(+1,-1)$ where the Greek
indices \,${\mu}\,,{\nu}\,,{\lambda}\,,... = 0\,,1$ correspond to the 
time and space directions, respectively. We choose 2D Levi-Civita tensor 
$\varepsilon_{\mu\nu}$ to obey the properties: $\varepsilon_{\mu\nu}\,
\varepsilon^{\mu\nu} =-\; 2!$, $\varepsilon_{\mu\nu}\,\varepsilon^{\nu\lambda} =\delta 
^{\lambda}_{\mu},\; \varepsilon_{01} =+1=\varepsilon^{10}$, etc. 
In 2D, the curvature tensor (i.e. field strength tensor)
$F_{\mu\nu}$ has only one existing component 
$F_{01} = E = -\varepsilon^{\mu\nu}[\partial_{\mu}A_{\nu} 
+\,\frac{i}{2}(A_{\mu}\times A_{\nu})]$ because 
$F_{\mu\nu} = \partial_{\mu}A_{\nu} -\partial_{\nu}A_{\mu}  
+ i\,(A_{\mu}\times A_{\nu})$. Here, in the $ SU(N)$
 Lie algebraic space, we have adopted
the notations $A\cdot B = A^a\,B^a$ and $(A\times B)^a = f^{abc}A^b\,B^c$ 
for the non-null vectors $A^a$ and $B^a$ where 
$a,b,c  = 1,2,.....,N^2-1$ and $f^{abc}$ are  the structure constants
 in the $SU(N)$ Lie algebra $[T^a,T^b] = f^{abc} T^c$ 
for the generators $T^a$ which are present in
the definition of $1$-form potential $A_{\mu}=A_{\mu}\cdot T=A_{\mu}^aT^a$ 
and  curvature $2$-form field strength tensor $F_{\mu\nu}=F_{\mu\nu}\cdot T=
F_{\mu\nu}^aT^a$, etc. Through out the whole body of our text, 
we denote the (anti-)BRST and (anti-)co-BRST fermionic ($s_{(a)b}^2 = s_{(a)d}^2 = 0$)
symmetry transformations by $s_{(a)b}$ and $s_{(a)d}$,
respectively.

\section
{\bf Preliminaries: Nilpotent (Fermionic) Symmetries}

We discuss here the (anti-)BRST and (anti-)co-BRST symmetries 
(and  derive their corresponding conserved charges) in the Lagrangian formulation of the 2D 
non-Abelian $1$-form $(A^{(1)} = dx^{\mu} A_{\mu} = dx^{\mu} A_{\mu}\cdot T$) 
gauge theory within the framework of BRST formalism. The starting coupled 
Lagrangian densities, in the Curci-Ferrari gauge [27,28], are:
\begin{eqnarray}
&&{\cal L}_B = {\cal B} {\cdot E} - \frac {1}{2}\,{\cal B} \cdot {\cal B} +\, B\cdot (\partial_{\mu}A^{\mu}) 
+ \frac{1}{2}(B\cdot B + \bar B \cdot \bar B) - i\,\partial_{\mu}\bar C \cdot D^{\mu}C, \nonumber\\
&&{\cal L}_{\bar B} = {\cal B} {\cdot E} - \frac {1}{2}\,{\cal B} \cdot {\cal B} - \bar B\cdot (\partial_{\mu}A^{\mu}) 
+ \frac{1}{2}(B\cdot B + \bar B \cdot \bar B) - i\, D_{\mu}\bar C \cdot \partial^{\mu}C,
\end{eqnarray}
where ${\cal B}$, $B$ and $\bar B $  are the Nakanishi-Lautrup type 
auxiliary fields that have been invoked for various purposes. For instance,
 ${\cal B}$ is introduced in the theory to linearize the kinetic term 
($ -\frac{1}{4}F_{\mu\nu}\cdot F^{\mu\nu}$ = $\frac{1}{2}E\cdot E \equiv {\cal B}\cdot E$ - $\frac{1}{2}
{{\cal B}}\cdot{{\cal B}}$) and auxiliary  fields $B$ and 
${\bar B} $ satisfy the Curci-Ferrari restriction:
 ${B + \bar B + (C\times \bar C)} = 0 $ where the 
(anti-)ghost fields ${\bar C}$ and $C$ are fermionic
  (i.e. $(C^a)^2= ({\bar C^a}) ^2=0, \; C^a \bar C^b+ \bar C^b C^a = 0,\;  C^a C^b+C^b C^a= 0,\; \bar C^a\bar C^b+\bar 
C^b\bar C^a = 0,\; \bar C^a C^b+C^b\bar C^a= 0,$ etc.) in nature and they are
 required in the theory for the validity of unitarity. 
In the above, we have the covariant derivatives 
$[D_{\mu} C = \partial_{\mu} C + i\, (A_{\mu}\times C) $ 
and $D_{\mu}\bar C = \partial_{\mu}\bar C + i\, (A_{\mu}\times\bar C)]$  on the (anti-)ghost fields in 
the adjoint representation.

The Lagrangian densities in (1) respect the following 
off-shell nilpotent $(s_{(a)b}^2 = 0)$ (anti-)BRST symmetries transformations ($s_{(a)b}$)
\begin{eqnarray}
&&s_b A_\mu = D_\mu C,  \;\,\,\,\,s_b C =  - \frac{i}{2} (C\times C),\,\,\,\;  s_b\bar C \;= i\,B ,\;
 \,\,\,\, \;s_b B = 0,\,\,\,\, \;s_b({\cal B}\cdot{\cal B}) = 0, \nonumber\\
&& s_b\bar B = i\,(\bar B\times C),\;\;\,\,\,s_b E = i\,(E\times C),\qquad \,\,\,s_b {\cal B} 
= i\,{(\cal B}\times C),\quad \,\,\,
s_b({\cal B}\cdot E) = 0,\nonumber\\
&&s_{ab} A_\mu= D_\mu\bar C,\,\,\,\,\,  s_{ab} \bar C= -\frac{i}{2}\,(\bar C\times\bar C),
\,\,\,\, s_{ab}C   = i{\bar B},\,\,\,\,  s_{ab}\bar B    = 0
,\,\,\,\, s_{ab}({\cal B}\cdot{\cal B}) = 0,\nonumber\\
&&s_{ab} E      = i \,(E\times\bar C),\,\;\;
\,\,\; s_{ab}{\cal B} =  i\,({\cal B}\times\bar C),
\,\,\, \; \;s_{ab} B = i\,(B \times \bar C),\quad \;s_{ab}({\cal B}\cdot E) = 0,              
\end{eqnarray}
because the Lagrangian densities ${\cal L}_B$ and ${\cal L}_{\bar B}$ transform under $s_{(a)b}$ as:
\begin{eqnarray}
&&s_b{\cal L}_B = \partial_\mu(B \cdot D^\mu C),  
\qquad\qquad\qquad\quad s_{ab}{\cal L}_{\bar B}= -\partial_\mu{(\bar B \cdot D^\mu \bar C)},\nonumber\\
&&s_{ab}{\cal L}_B = -\partial_\mu\,[{\{\bar B + ( C\times\bar C)\} \cdot \partial^\mu \bar C}\,] 
+\{B+\bar B + ( C \times {\bar C})\} \cdot D_\mu \partial^\mu \bar C, \nonumber\\
&&s_b{\cal L }_{\bar B}\; = \partial_\mu\,[ {\{ B + ( C \times \bar C )\}}\cdot
\partial^\mu C \,]-{\{B + \bar B + ( C\times\bar C )\}}\cdot D_\mu\partial^\mu C.
\end{eqnarray}
It should be noted that {\it both} the Lagrangian densities in Eq. (1)
respect {\it both} (i.e. BRST and anti-BRST) 
symmetries on the constrained hypersurface where the CF-condition $({B + \bar B + (C\times \bar C)} = 0) $ 
is satisfied. In other words, we note that $s_b{\cal L}_{\bar B} = - \partial_\mu[\bar B \cdot \partial^\mu C]$ and 
$s_{ab}{\cal L}_{B}= \partial_\mu{[ B \cdot \partial^\mu \bar C]}$ because of the validity of   
CF-condition. As a consequence, the action integrals ${S= \int d^2x \,{\cal L}_B}$ and
 ${S = \int d^2x \, {\cal L}_{\bar B}}$ remain invariant under the (anti-)BRST symmetries on the 
above hypersurface located in the 2D Minkowskian spacetime manifold. It is interesting  to point out that the absolute anticommutativity 
${\{s_b,s_{ab}}\} = 0$ is {\it also} satisfied on the above hypersurface which is defined by the field equation: 
${B + \bar B + (C\times \bar C)} = 0 $.

According to the celebrated Noether's theorem, the above continuous symmetries lead to the derivations of conserved currents and
charges. These (anti-)BRST charges, corresponding to the above continuous symmetries $s_{(a)b}$, are  (see, e.g. [14]  for details)
\begin{eqnarray}
&&Q_{ab} =\int dx\;[\dot{\bar B}\cdot\bar C -\bar B\cdot D_0\bar C +\frac{1}{2}(\bar C\times \bar C)\cdot\dot C],\nonumber\\
&&Q_b =\int dx \;[B\cdot D_0 C -\dot B\cdot C -\frac{1}{2}\,\dot{\bar C}\cdot(C\times C)],
\end{eqnarray}
where a single dot on a field denotes the ordinary time derivative (e.g. $\dot C = \partial \;C/\partial\; t$).

The above conserved charges $Q_{(a)b}$ are nilpotent $(Q_b^2 = Q_{ab}^2 = 0)$ of order two and they obey absolute anticommutativity property (i.e. $Q_b Q_{ab} + Q_{ab}Q_b = 0$).
These properties can be mathematically expressed as follows:
\begin{eqnarray}
&&s_b Q_b = -i\,\,{\{Q_b,Q_b}\} = 0\Longrightarrow \quad Q_b^2 = 0,\quad\,
s_{ab} \,Q_{ab} = -i\,\{Q_{ab},Q_{ab}\} = 0 \Longrightarrow \,\, Q_{ab}^2 = 0,
\nonumber\\
&&s_{ab}Q_{b} = -i\,{\{Q_{b},Q_{ab}}\} = 0\Longrightarrow \{Q_{b},Q_{ab}\} = 0 
\quad \Longleftrightarrow\;\;  \,s_b Q_{ab}  = -i\,{\{Q_{ab},Q_b}\} = 0.
\end{eqnarray}
The preciseness of the above expressions can be verified by taking into account the 
nilpotent (anti-)BRST symmetry transformation $s_{(a)b}$ (cf. Eq. (2))
and expressions for the nilpotent (anti-)BRST  charges from Eq. (4). 
It should be noted that the property of absolute anticommutativity of 
the (anti-)BRST charges (i.e. ${\{Q_b,Q_{ab}}\} = 0$)
is {\it true} only when we use the CF-condition (i.e.  ~${B + \bar B + (C\times \bar C)} = 0) $.

The Lagrangian densities (1) {\it also} respect the following off-shell
 nilpotent ($ \,s_{(a)d}^2 = 0$) and absolutely anticommuting
 ($s_ds_{ad} + s_{ad}s_d = 0$) (anti-)co-BRST [i.e. (anti-)dual BRST] 
symmetry transformations $(s_{(a)d})$ (see, e.g. [13], [14])
\begin{eqnarray}
&&s_{ad} A_{\mu} = - \varepsilon_{\mu\nu}\partial^\nu C,\quad\quad s_{ad} C = 0,\quad
\quad s_{ad} \bar C = i\; {\cal B},\quad s_{ad} B = 0, \nonumber\\
&&s_{ad}\bar B = 0,\quad s_{ad} E = D_\mu\partial^\mu C,\quad\quad s_{ad}({\partial_\mu A^\mu})= 0,
\quad s_{ad} {\cal B} = 0, \nonumber\\
&&s_d A_\mu = - \varepsilon_{\mu\nu} \partial^\nu \bar C,
\qquad s_d \bar C = 0,\quad\quad\quad s_d C = - i\; {\cal B},\quad s_d B = 0,\nonumber\\
&&s_d{\bar B} = 0,\quad s_d E = D_\mu\partial^\mu\bar C,\qquad s_d({\partial_\mu A^\mu})= 0,\quad\quad s_d{\cal B} = 0,
\end{eqnarray}
because  the above Lagrangian densities transform, under $s_{(a)d}$, as follows:
\begin{eqnarray}
&&s_{ad}\,{\cal L}_{\bar B} = \partial_\mu[{\cal B}\cdot \partial^\mu C],  \qquad\qquad\qquad\quad s_d {\cal L}_B = \partial_\mu[{\cal B}
\cdot\partial^{\mu} \bar C],\nonumber\\ 
&& s_{ad}{\cal L}_B = \partial_\mu[{\cal B}\cdot D^\mu C+ \varepsilon^{\mu\nu}\bar C\cdot(\partial_\nu C\times C)] 
+ i\; (\partial_\mu A^\mu)\cdot({\cal B}\times C),\nonumber\\&& s_d{\cal L}_{\bar B} = 
\partial_\mu[{\cal B}\cdot D^\mu\bar C- \varepsilon^{\mu\nu}C \cdot(\partial_\nu\bar C\times  
\bar C)] + i\; (\partial_\mu A^\mu)\cdot({\cal B}\times\bar C).
\end{eqnarray}
It is clear that {\it both} the Lagrangian densities respect {\it both} (i.e. co-BRST and anti-co-BRST) 
fermionic symmetry transformations on a hypersurface where the 
CF-type restrictions ${\cal B}\times C =0 $, ${\cal B}\times\bar C=0 $ are satisfied. We lay emphasis on the observation 
that absolute anticommutativity $\{s_{d}, s_{ad}\} = 0$ is satisfied {\it without} any use of CF-type restrictions
${\cal B}\times C =0 $, ${\cal B}\times\bar C=0 $. More 
elaborate discussions about these CF-type restrictions (and other related restrictions) can be found in our 
earlier works (see, e.g., [14,16] for details).

The Noether conserved ($\dot Q_{(a)d} = 0$) charges $Q_{(a)d}$, corresponding to the continuous and nilpotent symmetry transformations (6), are:
\begin{eqnarray}
&&Q_d =\int dx\;[{\cal B}\cdot\dot{\bar C}+B\cdot\partial_1\bar C]
\;\;\equiv  \int dx\;[{\cal B}\cdot\dot{\bar C}-D_0{\cal B}\cdot\bar C 
+(\partial_1\bar C\times C)\cdot\bar C],\nonumber\\
&&Q_{ad}=\int dx\;[{\cal B}\cdot\dot C- \bar B \cdot\partial_1 C] 
\equiv \int dx\;[{\cal B}\cdot\dot C-D_0{\cal B}\cdot C -(\bar C\times \partial_1 C)\cdot C].
\end{eqnarray}
The above charges are found to be nilpotent $(Q_{(a)d}^2 = 0)$ and absolutely anticommuting $(Q_dQ_{ad} + Q_{ad}Q_d = 0)$ 
in nature. These claims can be verified in a straightforward fashion by taking the help of symmetries (6) and expressions of the charges (8)
as follows:
\begin{eqnarray}
&&s_d Q_d = -i\,\,{\{Q_d,Q_d}\} = 0\Longrightarrow \quad Q_d^2 = 0,\quad s_{ad} \,Q_{ad} = -i\,\{Q_{ad},Q_{ad}\} = 0 \Longrightarrow \,\, Q_{ad}^2 = 0,
\nonumber\\
&&s_d Q_{ad} = -i\,{\{Q_{ad},Q_d}\} = 0\quad\Longleftrightarrow \,\, \,s_{ad} Q_d  = -i\,{\{Q_{d},Q_{ad}}\} = 0~\Longrightarrow~ {\{Q_d, Q_{ad}}\} = 0.
\end{eqnarray}
In fact, in this simple proof, one has to verify the  l.h.s. of the above equations. In the forthcoming sections, we shall exploit the beauty 
and strength of the AVSA to BRST formalism to capture the above properties in a cogent and consistent manner.

\section{Horizontality Condition: Off-Shell Nilpotent\\ (Anti-)BRST Symmetry Transformations}

We concisely mention here the key points associated with the geometrical origin of the nilpotent (anti-)BRST symmetries and existence
of the CF-condition within the framework of Banora-Tonin (BT) superfield formalism [4,5]. In this connection, first of all,
we generalize the 2D ordinary theory onto $(2,2)$-dimensional supermanifold where the non-Abelian $1$-form gauge field $A_{\mu}(x)$
and (anti-)ghost fields $({\bar C}) C$ are generalized onto their  corresponding superfields with the following expansions 
(incorporating the secondary fields $R_\mu, \bar R_\mu, S_\mu,B_1, B_2, \bar B_1, \bar B_2, s, \bar s$)
on the 
$(2,2)$-dimensional supermanifolds  [4,5]:
\begin{eqnarray}
&&A_\mu(x)\to B_\mu (x,\theta,\bar\theta) = A_\mu (x) + \theta \,\bar R_\mu (x)
+ \bar \theta \, R_\mu (x) + i\, \theta \bar \theta \,S_\mu (x), \nonumber \\
&&C(x)\to F(x,\theta,\bar\theta) = C(x) + i\, \theta \,\bar B_1 + i\, \bar\theta \,B_1 + i\, \theta \bar \theta \,s(x), \nonumber \\
&&\bar C(x) \to \bar F(x,\theta,\bar\theta) = \bar C(x) + i \,\theta\, \bar B_2 +  i\, \bar \theta \, B_2 
+ i \,\theta \bar \theta \, \bar s(x),
\end{eqnarray}
where the supermanifold is characterized by the superspace coordinates $Z^M = (x^\mu,\theta,\bar\theta)$.
The 2D ordinary  bosonic coordinates $x^\mu\;(\mu = 0, 1)$ and the Grassmannian coordinates $(\theta, \bar\theta)$
(with $\theta^2 = \bar\theta^2 = \theta\bar\theta + \bar\theta\theta = 0)$ specify the superspace coordinate $Z^M$ and 
all the superfields, defined on the
supermanifold, are function of them. The super curvature $2$-form is
\begin{eqnarray}
\tilde F^{(2)} = \Bigl(\frac{dZ^M\,\wedge dZ^N}{2!}\Bigr)\,\tilde F_{MN}(x,\theta,\bar\theta) \quad\equiv  
\quad\tilde d \tilde A^{(1)} + i\,(\tilde A^{(1)}\wedge \tilde A^{(1)}),
\end{eqnarray}
where the super curvature tensor
 $\tilde F_{MN} = (\tilde F_{\mu\nu},\tilde F_{\mu\theta},\tilde F_{\mu\bar\theta}, 
\tilde F_{\theta\theta}, \tilde F_{\theta \bar\theta}, \tilde F_{\bar\theta \bar\theta})$.  In the above equation,
 the ordinary exterior derivative  $d = dx^\mu\,\partial_\mu$ 
and non-Abelian 1-form $(A^{(1)} = dx^\mu\,A_\mu)$ gauge connection have been generalized onto the (2, 2)-dimensional supermanifold as
\begin{eqnarray}
&&d = dx^\mu \partial_\mu \to \tilde d= dx^\mu\,\partial_\mu + d\theta\,\partial_\theta + d\bar\theta \,\partial_{\bar\theta}, 
\quad \qquad\quad \tilde d^2 = 0,\nonumber\\
&&A^{(1)} = dx^\mu A_\mu \to \tilde A^{(1)} = dx^{\mu}\,B_{\mu}(x, \theta, \bar\theta) + d\theta\,\bar F(x, \theta, \bar\theta)
+ d\bar\theta\,F(x, \theta, \bar\theta),
\end{eqnarray}
where $ (\partial_\mu,\partial_\theta,\partial_{\bar\theta})$ are the superspace derivatives (with
$\partial_\mu = \frac{\partial}{\partial x^\mu}$, $\partial_\theta = \frac{\partial}{\partial \theta}$ and $\partial_{\bar\theta} = 
\frac\partial{\partial \bar\theta}$).

We have observed earlier that the kinetic term $(-\frac{1}{4}\,F_{\mu\nu}\cdot F^{\mu\nu} = {\cal B}\cdot E - \frac{{\cal B}\cdot{\cal B}}{2})$
of the Lagrangian densities (1) remains invariant under the (anti-)BRST symmetries (2) and it has its origin in the exterior derivative 
$d$ (i.e. $F^{(2)}=\,dA^{(1)}+ i\,A^{(1)}\wedge A^{(1)}$). This gauge invariant quantity should remain independent of the 
Grassmannian variables $(\theta, \bar\theta)$ as the latter are only a {\it mathematical} artifacts and they cannot be physically 
realized. Thus, we have the following equality due to the gauge invariant restriction (GIR):
\begin{eqnarray}
-\frac{1}{4}\,\tilde F_{MN}(x,\theta,\bar\theta)\cdot \tilde F^{MN}(x,\theta,\bar\theta) 
= -\frac{1}{4}\,F_{\mu\nu}(x)\cdot F^{\mu\nu}(x).
\end{eqnarray}
The celebrated horizontality condition (HC) requires that the Grassmannian components of 
$\tilde F_{MN}\;(x,\theta,\bar\theta) = (\tilde F_{\mu\nu},\tilde F_{\mu\theta},\tilde F_{\mu\bar\theta}, \tilde F_{\theta\theta}, \tilde F_{\theta \bar\theta}, \tilde F_{\bar\theta \bar\theta})$
should be set equal to zero 
so that, ultimately, we should have the following equality, namely;
\begin{eqnarray}
&&-\frac{1}{4}\,\tilde F_{\mu\nu}(x,\theta,\bar\theta)\cdot \tilde F^{\mu\nu}(x,\theta,\bar\theta) 
= -\frac{1}{4}\;F_{\mu\nu}(x)\cdot F^{\mu\nu}(x).
\end{eqnarray}
The requirement of HC leads to the following [4,5,11,12]
\begin{eqnarray}
&&R_{\mu}(x) = D_{\mu}C,\qquad \bar R_{\mu}(x) = D_{\mu}\bar C,\qquad s = i\;(\bar B\times C),\quad\bar s = -i\,(B\times \bar C),\nonumber\\
&&S_\mu  = (D_\mu B + D_\mu C\times \bar C)\equiv  -(D_\mu \bar B + C\times D_\mu \bar C),\quad B_1 = -\frac{1}{2}\,(C \times C), \nonumber \\ 
&& \bar B_2 = -\frac {1}{2}\,(\bar C\times \bar C),\quad B_1 + B_2 + (C\times\bar C) = 0, 
\end{eqnarray}
where the last entry is nothing but the celebrated CF-condition $(B + \bar B + (C\times \bar C) = 0)$ if we identify $\bar B_1 = \bar B$
and $B_2 = B$. It is crystal clear that the HC leads to the derivation of the secondary fields in terms of the
auxiliary and basic fields of the starting Lagrangian densities (1). The substitution of the above expressions
for the secondary  fields into the super expansion (10)
leads to the following [4,5,11,12]
\begin{eqnarray}
B_{\mu}^{(h)}(x, \theta, \bar\theta) &= & A_{\mu}(x) + \theta \, (D_{\mu}\bar C) + \bar\theta \,(D_\mu C) + 
i\,\theta\bar\theta \, [D_{\mu}B + D_\mu C\times \bar C]  \nonumber\\
&\equiv & A_{\mu}(x) + \theta \, (s_{ab}A_{\mu}) + \bar\theta \,(s_{b}A_{\mu}) + \theta\bar\theta \,(s_{b}s_{ab}A_{\mu}), 
 \nonumber\\
F^{(h)}(x, \theta, \bar\theta) &= & C(x) + \theta \,(i \bar B) + \bar\theta \, [-\frac{i}{2}(C\times C)] + 
\theta\bar\theta \, (- \bar B\times  C)  \nonumber\\
&\equiv & C(x) + \theta \, (s_{ab} C) + \bar\theta \,(s_{b} C) + \theta\bar\theta \,(s_{b}s_{ab} C),  \nonumber\\
\bar F^{(h)}(x, \theta, \bar\theta) &= & \bar C(x)  
+ \theta \,[-\frac{i}{2}(\bar C\times \bar C)] + \bar\theta \,(i B)
+ \theta\bar\theta \,(B\times \bar C)  \nonumber\\
&\equiv & \bar C(x) + \theta \,(s_{ab} \bar C) + \bar\theta \,(s_{b} \bar C) + \theta\bar\theta \,(s_{b}s_{ab}\bar C),
\end{eqnarray}  
where the superscript $(h)$ on the superfields denotes the fact that these superfields have been obtained after the 
application of  HC. A close look at the above expressions demonstrate that the coefficients of $(\theta,\bar\theta)$
are nothing but the anti-BRST and BRST transformations (2), respectively, that have been listed for the Lagrangian densities (1).

Due to application of HC, ultimately, we obtain the following expression for the supercurvature tensor
(as we have already set $\tilde F_{\mu\theta}=\tilde F_{\mu\bar\theta}= \tilde F_{\theta \bar\theta}=\tilde F_{\theta\theta}
 = \tilde F_{\bar\theta \bar\theta} = 0)$: 
\begin{eqnarray} 
&&\tilde F_{\mu\nu}^{(h)}(x, \theta, \bar\theta)  = \partial_\mu\, B_\nu^{(h)} - \partial_\nu\, B_\mu^{(h)} 
+ i\,(B_\mu^{(h)}\times  B_\nu^{(h)}).    
\end{eqnarray}
Substitution of the expression for $B_{\mu}^{(h)}(x, \theta, \bar\theta)$, from (16), yields 
\begin{eqnarray} 
\tilde F_{\mu\nu}^{(h)}(x, \theta, \bar\theta) &=&  F_{\mu\nu}(x) + \theta\,(i\,F_{\mu\nu}\times \bar C) +  \bar \theta\,(i\,F_{\mu\nu}\times  C)\nonumber\\  
&+&\theta\bar\theta \;[-(F_{\mu\nu}\times C)\times \bar C - F_{\mu\nu}\times B]\nonumber\\
&\equiv & F_{\mu\nu}^ {(h)} + \theta\;(s_{ab} F_{\mu\nu}) + \bar\theta\;(s_{b} F_{\mu\nu}) + \theta\bar\theta\;(s_{b}s_{ab}F_{\mu\nu}),  
\end{eqnarray}
which leads to the derivation of the (anti-)BRST symmetry 
transformations for the $F_{\mu\nu}$ (cf. (2)). It is now crystal clear  that the 
requirements of gauge invariant restrictions in (14) and 
(13) are satisfied due to HC and, in this process, we have
 obtained the (anti-)BRST symmetry transformations 
for {\it all} the fields (as well as the CF-condition)
for our theory. We have derived these (anti-)BRST symmetry
transformations by exploiting  the potential of (anti-)chiral
superfields approach to BRST formalism in our Appendix A.

\noindent
\section {Dual Horizontality Condition: Nilpotent (Anti-)co-BRST Symmetry Transformations}

We exploit here the dual-HC (DHC) to derive the (anti-)co-BRST symmetry transformations
for the (anti-)ghost fields and basic tenets of AVSA to obtain the precise form
of the (anti-)co-BRST symmetry transformations associated with the gauge field
$(A_{\mu} = A_{\mu}\cdot T)$ of our 2D non-Abelian theory.
In this context, first of all, we note that the gauge-fixing term $(\partial_{\mu} A^{\mu})$ 
 has its origin in the co-exterior derivative $(\delta  = - \ast\;d\;\ast)$ of the differential geometry in the 
 following sense (see, e.g. [21-24] for details)
\begin{eqnarray}
&& \delta A ^{(1)} = -\ast\; d\;\ast  (dx^{\mu}A_{\mu}) = \partial_{\mu}A^{\mu},\qquad\qquad       \delta^2 = 0,
\end{eqnarray}
where $ \delta = -\ast\; d\;\ast $ is the co-exterior derivative and $ \ast $ is the Hodge duality operator on 2D Minkowskian {\it flat} spacetime 
manifold. It is clear that the Lorentz gauge-fixing term $(\partial_{\mu}A^\mu)$ is a $0$-form which emerges out from the 1-form
$(A^{(1)} = dx^\mu A_\mu)$ due to application of the co-exterior derivative $(\delta = -\ast\; d\;\ast)$ which reduces the degree of 
 a form by one.

We have seen that the gauge-fixing term $(\partial_{\mu}A^\mu)$ remains invariant under the (anti-)co-BRST symmetry transformations
(cf. Eq. (6)). We generalize this observation onto our chosen (2, 2)-dimensional supermanifold as follows
\begin{eqnarray}
&& \tilde\delta \tilde A ^{(1)} = \delta A^{(1)},\qquad\qquad \tilde\delta =  -\star\; \tilde d\;\star  ,\qquad\qquad    \tilde\delta^2 = 0\qquad\qquad 
\tilde d ^ 2 = 0,
\end{eqnarray}
where $\tilde\delta$ is the super co-exterior derivative defined on the (2, 2)-dimensional supermanifold and $\star$ is the Hodge duality 
operator on the (2, 2)-dimensional supermanifold (see, e.g. [29] for details). The l.h.s. of (20) has already been computed 
in our previous work [29]. We quote here the result of operation of $\tilde\delta$ on $ \tilde A ^{(1)} $
as 0-form, namely;
\begin{eqnarray}
\partial_{\mu} B^\mu + \partial_\theta \bar F + \partial_{\bar\theta} F + s^{\bar\theta\bar\theta}(\partial_{\bar\theta}\bar F) + 
s^{\theta\theta}(\partial_\theta F) = \partial_\mu A^\mu,
\end{eqnarray}
where $s^{\theta\theta}$ and $s^{\bar\theta\bar\theta}$ appear in the following Hodge duality $\star$ operation:
\begin{eqnarray}
\star\;(dx_\mu \wedge dx_\nu \wedge d\bar\theta \wedge d\bar\theta) = \varepsilon_{\mu\nu}\; s^{\bar\theta\bar\theta}\nonumber\\
\star\;(dx_\mu \wedge dx_\nu \wedge d\theta \wedge d\theta) = \varepsilon_{\mu\nu}\; s^{\theta\theta}.
\end{eqnarray} 
These factors (i.e. $s^{\theta\theta}$, $s^{\bar\theta\bar\theta}$) are essential to get back the 4-forms 
$(dx_\mu \wedge dx_\nu \wedge d\theta \wedge d\theta)$ and
$(dx_\mu \wedge dx_\nu \wedge d\bar\theta \wedge d\bar\theta)$ if we apply {\it another} $\star$ on (22). 
In other words, we have super Hodge duality $\star$ on the $0$-form as follows:
\begin{eqnarray}
\star\;(\varepsilon_{\mu\nu} s^{\theta\theta}) \;= \;\pm \;(dx_\mu \wedge dx_\nu \wedge d\theta \wedge d\theta),\nonumber\\
\star\;(\varepsilon_{\mu\nu} s^{\bar\theta\bar\theta})\; = \;\pm \;(dx_\mu \wedge dx_\nu \wedge d\bar\theta \wedge d\bar\theta).
\end{eqnarray}
The equality in (21) ultimately, leads to 
\begin{eqnarray}
\partial_\theta F = 0,\qquad\qquad \partial_{\bar\theta}\bar F = 0,\qquad\qquad\partial_{\mu} B^\mu + 
\partial_\theta \bar F + \partial_{\bar\theta} F = \partial_\mu A^\mu,
\end{eqnarray}
because of the fact that there are {\it no} terms carrying the factors $s^{\theta\theta}$ and $s^{\bar\theta\bar\theta}$ on the r.h.s.

At this stage, we substitute the expressions of $B_\mu (x, \theta, \bar\theta)$, $F(x, \theta, \bar\theta)$ and $\bar F(x, \theta, \bar\theta)$
into Eq. (24) to derive the following important relationships:
\begin{eqnarray}
&&\partial_\mu R^\mu = 0,\qquad \partial_\mu \bar R^\mu = 0, \qquad\qquad \partial_\mu S^\mu = 0, \qquad\qquad  s = 0,\nonumber\\
&&\bar B_1 = 0,\qquad\qquad B_2 = 0,\qquad\qquad \bar s = 0,\qquad\qquad B_1 + \bar B_2 = 0.
\end{eqnarray}
The last entry, in the above, is just like the CF-type restriction which is {\it trivial}. With the choices $B_1 = - \cal B$ and 
$\bar B_2 = \cal B$, we obtain the following expansions
\begin{eqnarray}
&&F^{(dh)}(x,\theta,\bar\theta) = C(x) + \bar\theta\; (-\;i{\cal B})\equiv C(x) + \bar\theta\; (s_d C),\nonumber\\
&&\bar F^{(dh)}(x,\theta,\bar\theta) = \bar C(x) + \theta \;(\;i{\cal B})\equiv \bar C(x) + \bar\theta \;(s_{ad}\bar C),
\end{eqnarray} 
where the superscript $(dh)$ denotes the  expansions of the superfields after the 
application of DHC. It is self-evident that we have already obtained the (anti-)co-BRST symmetry
transformation (4) for the (anti-)ghost fields $(\bar C)C$ of our theory as:
\begin{eqnarray}
&& s_d C = - i\; {\cal B},\qquad s_{ad} C = 0,\qquad s_d\bar C = 0,\qquad s_{ad} C =  i~ {\cal B}.
\end{eqnarray}
Thus, the DHC leads to the derivation of (anti-)co-BRST symmetry transformations for the 
(anti-)ghost fields and very useful restrictions on the secondary fields in (25).

We are now in the position to derive the (anti-)co-BRST  symmetry transformations $s_{(a)d}$
for the gauge field $A_{\mu}$. We exploit here the idea of AVSA to BRST formalism which states that
the (anti-)co-BRST invariant quantities should be independent of the ``soul'' coordinates $(\theta,\bar\theta)$. During the early
days of the developments of superspace technique, the bosonic coordinates $x^\mu$ of the superspace coordinates
$Z^M = (x^\mu, \theta, \bar\theta)$ were called as the ``body'' coordinates and the Grassmannian variables
$(\theta, \bar\theta)$ were christened as the ``soul'' coordinates.
In this context, we observe that the following is true, namely;
\begin{eqnarray}
&& s_{(a)d}\; [ \varepsilon^{\mu\nu} A_{\nu}\cdot \partial_{\mu} {\cal B}
- i\; \partial_{\mu}\bar C\cdot \partial^{\mu} C] = 0.
 \end{eqnarray} 
Thus, we have the following equality  due to AVSA to BRST formalism:
\begin{eqnarray}
&& \varepsilon^{\mu\nu} B_{\nu}(x,\theta,\bar\theta)\cdot\partial_{\mu} {\cal B}(x) - i\; \partial_{\mu} \bar F^{(dh)}(x,\theta,\bar\theta)\cdot
\partial^{\mu} F^{(dh)}(x,\theta,\bar\theta)\nonumber\\
&& \equiv  \varepsilon^{\mu\nu} A_{\nu}(x)\cdot\partial_{\mu}{\cal B}(x) -
i\;\partial_{\mu}\bar C (x) \cdot\partial^{\mu} C(x).
\end{eqnarray}
The substitution of the expansions from (26) yields the following:
\begin{eqnarray}
&&\varepsilon^{\mu\nu} \bar R_\nu + \partial^\mu C = 0, \qquad\qquad \varepsilon^{\mu\nu} R_\nu + \partial^\mu \bar C = 0,
\qquad\qquad \varepsilon^{\mu\nu}S_\nu - \partial^\mu{\cal B} = 0.
\end{eqnarray}
It is worthwhile to point out that we have {\it not} taken any super expansion of ${\cal B}(x)$ on the l.h.s. in (29) because of the fact that 
$s_{(a)d} {\cal B} (x) = 0$. In other words, we have taken ${\cal B}(x)\longrightarrow \tilde{\cal B}(x,\theta,\bar\theta) = {\cal B}(x)$. 
Ultimately, the relation in (30) produces the following:
\begin{eqnarray}
R_\mu = -\;\varepsilon_{\mu\nu} \partial^{\nu}\bar C, \qquad\qquad \bar R_\mu = -\;\varepsilon_{\mu\nu} \partial^{\nu}C,\qquad\qquad
S_\mu = \varepsilon_{\mu\nu} \partial^{\nu}{\cal B}.
\end{eqnarray}
The substitution of these expressions into the super expansions of $B_{\mu} ( x, \theta, \bar\theta)$
leads to the following (in terms of the (anti-)co-BRST symmetry transformations (6)):
\begin{eqnarray}
B_{\mu}^{(dg)}( x, \theta, \bar\theta) & = & A_{\mu}(x) + \theta \;(\varepsilon_{\mu\nu} \partial^{\nu}C) +
 \bar\theta\; (-\;\varepsilon_{\mu\nu} \partial^{\nu}\bar C) + \theta\;\bar\theta\;+(\varepsilon_{\mu\nu} \partial^{\nu}{\cal B})\nonumber\\
 &\equiv & A_{\mu}(x) + \theta\;(s_{ad} A_\mu) + \bar\theta\;(s_d A_\mu) + \theta\;\bar\theta\;(s_d s_{ad}A_\mu).
\end{eqnarray}
Here the superscript $(dg)$ on $B_{\mu} ( x, \theta, \bar\theta)$ denotes 
the expansion that has been obtained after the application of 
(anti-)co-BRST (i.e. dual gauge) invariant restriction (29). We end 
this section with the remark that we have obtained {\it all}
the (anti-)co-BRST symmetry transformations for our 2D 
non-Abelian 1-form gauge theory by exploiting the theoretical strength of DHC and
basic tenets of AVSA
to BRST formalism.

\section{Nilpotency and Absolute Anticommutativity of the Fermionic Charges: Ordinary 2D Spacetime}

We, first of all, capture the nilpotency and absolute anticommutativity of the (anti-)BRST
 and (anti-)co-BRST charges in the {\it ordinary} space
where the concepts/ideas behind the continuous 
symmetry and their generators (as well as the nilpotency of the (anti-)BRST and (anti-)co-BRST
symmetry transformations) play very important roles. We would like to lay stress
 on the fact that some of the key results of our present section have been obtained due to
 our knowledge of the AVSA to BRST formalism that is contained in Sec. 6. Towards this goal in mind, we observe 
a few aspects of  the conserved charges (listed in Eq. (4) and Eq. (8)) corresponding 
to the (anti-)BRST and (anti-)co-BRST symmetries of the Lagrangian  densities (1).
Using the (anti-)BRST and (anti-)co-BRST symmetry transformations of Eq. (2) and Eq. (6), 
we observe that the following are true, namely;
\begin{eqnarray}
&&Q_b = s_b\;\Big(\int\; dx\; [ B\cdot A_0  + i\;\dot {\bar C}\cdot C]\Big),\qquad 
Q_{ab} = s_{ab}\;\Big(\int\; dx\; [ i\; \bar C\cdot\dot C - \bar B\cdot A_0]\Big),\nonumber\\
&&Q_d = s_d\;\Big(\int\; dx\; [ {\cal B}\;\cdot A_1  +  B\cdot A_0]\Big),\qquad Q_{ad} = s_{ad}\;\Big(\int\;dx\;[{\cal B}\cdot A_1 - \bar B\cdot A_0]\Big).
\end{eqnarray}
It should be noted that we have expressed the conserved charges in (4) and (8) in terms of 
the transformations in Eqs. (2) and (6). 
It is elementary now to check that the nilpotency of the charges is satisfied
\begin{eqnarray}
&&s_b Q_b = - i\,{\{Q_b, Q_b}\} =  0,\qquad\qquad s_{ab} Q_{ab} = - i\; {\{Q_{ab}, Q_{ab}}\} = 0,\nonumber\\
&&s_d Q_d = - i\,{\{Q_d, Q_d}\} =  0,\qquad\qquad s_{ad} Q_{ad} = - i\; {\{Q_{ad}, Q_{ad}}\} = 0,
\end{eqnarray}
due to the nilpotency  properties of (anti-)BRST and (anti-)co-BRST symmetry transformations
(i.e. $s_{(a)b}^ 2 = 0, s_{(a)d}^ 2 = 0$). In the above, we have used the basic principles behind 
the continuous symmetries and symmetry generators (as the conserved charges of the theory).
We {\it also} point out that we have taken into account {\it one} of the expressions for $Q_{(a)b}$
and $Q_{(a)d}$ from Eq. (4) and Eq. (8) that have been explicitly derived in Sec. 2.

To prove the absolute anticommutativity properties of the (anti-)BRST and 
(anti-) co-BRST conserved charges, we note the following useful relationships
\begin{eqnarray}
&&Q_d = s_{ad}\Big[\int\; dx\; \big(- i\;\bar C\cdot\dot{\bar C} + \frac {\bar C}{2}\cdot (A_0\times\bar C)\big)\Big],\nonumber\\
&&Q_{ad}= s_d\Big[\int\; dx \;\big(i\; C\cdot\dot C - \frac { C}{2}\cdot (A_0\times C)\big)\Big],\nonumber\\
&&Q_b = s_{ab}\Big[\int\; dx \;\big(i\; C\cdot\dot C - \frac { C}{2}\cdot (A_0\times C)\big)\Big],\nonumber\\
&&Q_{ab} = s_b\Big[\int\; dx\; \big(-i\;\bar C\cdot\dot{\bar C} +\frac {\bar C}{2}\cdot (A_0\times\bar C)\big)\Big],
\end{eqnarray}
which establish the absolute anticommutativity properties of 
the (anti-)co-BRST and  nilpotent (anti-)BRST charges  as follows
\begin{eqnarray}
&&s_{ad}\;Q_d = - i\; {\{Q_d, Q_{ad}}\} = 0,\qquad s_d\; Q_{ad} = - i\; {\{Q_{ad}, Q_d}\} = 0,\nonumber\\
&&s_{ab}\;Q_b = - i\; {\{Q_b, Q_{ab}}\} = 0,\qquad s_b \;Q_{ab} = - i\; {\{Q_{ab}, Q_b}\} = 0,
\end{eqnarray}
due to, once again, the nilpotency ($s_{(a)b}^ 2 = 0, s_{(a)d}^ 2 = 0$) properties of the (anti-)BRST
and (anti-)co-BRST symmetry transformations. It is 
interesting to point out that the expressions in the square brackets for the pair $(Q_b, Q_{ad})$
and the pair $(Q_d,Q_{ab})$ are {\it exactly} the {\it same} (as is evident from Eq. (35)). We would like to make a few remarks at this stage.
A close look at equations (35) and (36) establishes one of the key observations that the nilpotency 
of  symmetries {\it and} absolute anticommutativity properties of the conserved (anti-)BRST and (anti-)co-BRST
charges are inter-related. Furthermore, we would like to mention that, in
the expressions for $Q_{(a)d}$ in (35), we have dropped total space derivative terms in our computations.
It is very important to emphasize here that in the expressions for $Q_{(a)b}$ (cf. Eq. (4)), we have utilized the strength
of CF-condition $(B + \bar B +  (C\times \bar C) = 0)$ to recast these expressions in a {\it suitable} form before expressing 
them in the form (35). To elaborate on it, we take a simple example where the expression for the BRST charges $Q_b$
(cf. Eq. (4)), emerging from the Noether conserved current, is:
\begin{eqnarray}
&&Q_b = \int\; dx\;\Big[B\cdot D_0 C - \dot B\cdot C - \frac{\dot {\bar C}}{2}\cdot\;(C\times C)\Big].
\end{eqnarray}
 Using the CF-condition $B + \bar B +  (C\times \bar C) = 0$ (associated with the (anti-)BRST symmetries),
 we can recast the above  expression in the following suitable
form:
\begin{eqnarray}
Q_b &=& \int\; dx\; \Big[\dot{\bar B}\cdot C - \bar B\cdot D_0 C - (C\times\bar C)\cdot D_0 C
+ \frac {\dot{\bar C}}{2}\cdot (C\times C) + (\dot C \times\bar C)\cdot C\Big], \nonumber\\
&\equiv & \int\; dx\;\Big[\dot{\bar B}\cdot C - \bar B\cdot D_0 C + \frac{\dot {\bar C}}{2}\cdot\;(C\times C) - i\,(C \times\bar C)\cdot (A_0 \times C)\Big ]. 
\end{eqnarray} 
The above form of the BRST charge has been expressed in the anti-BRST {\it exact} form as given in Eq. (35).
 A similar kind of argument has gone into the expression for the anti-BRST charge $Q_{ab}$ (cf. (35)) where
 we have been able to express it as the BRST {\it exact} form.
 No such kinds of arguments have been invoked in the cases of the (anti-)co-BRST charges (cf. (35))
which have been expressed as the co-BRST {\it exact} and anti-co-BRST {\it exact} forms.

We have modified the Lagrangian densities (1) in our earlier works [14,16] by 
incorporating a couple of fermionic Lagrange multiplier fields
 $(\lambda,\bar\lambda$ with $\lambda^2 =\bar\lambda^2 = 0$, $\lambda\bar\lambda + \bar\lambda\lambda = 0$)
in such a manner that the following modified Lagrangian densities [14,16]
\begin{eqnarray}
 {\cal L}^{(\bar\lambda)}_B &=& {\cal B}{\cdot E}-\frac {1}{2}\,{\cal B} \cdot {\cal B} 
+ B\cdot (\partial_{\mu}A^{\mu}) + \frac{1}{2}(B\cdot B 
+ \bar B \cdot\bar B)\nonumber\\
&-& i\,\partial_{\mu}\bar C \cdot D^{\mu}C + \bar\lambda\cdot({\cal B}\times C),\nonumber\\
{\cal L}^{(\lambda)}_{\bar B} &=& {\cal B} {\cdot E}-\frac {1}{2}\,{\cal B} \cdot {\cal B} - \bar B\cdot (\partial_{\mu}A^{\mu}) 
+ \frac{1}{2}(B\cdot B + \bar B \cdot \bar B)\nonumber\\
&-& i\, D_{\mu}\bar C \cdot \partial^{\mu}C +
\lambda\cdot({\cal B}\times\bar C),
\end{eqnarray}
respect the following {\it perfect} (anti-)co-BRST symmetries transformations:
\begin{eqnarray}
&&s_{ad} A_{\mu} = - \varepsilon_{\mu\nu}\partial^\nu C,\quad\quad s_{ad} C = 0,\quad
\quad s_{ad}\,  \bar C =  i\; {\cal B},\qquad\quad s_{ad} {\cal B} = 0,\nonumber\\
&&s_{ad} E =D_\mu\partial^\mu C,\quad\quad s_{ad}({\partial_\mu A^\mu})= 0, \quad 
s_{ad}\,{\lambda}= -i\;({\partial_\mu A^\mu}),\quad s_{ad}\,{\bar \lambda}= 0, \nonumber\\
&&s_d A_\mu = - \varepsilon_{\mu\nu}\partial^\nu \bar C,
\quad\quad s_d \bar C = 0,\quad\quad\quad s_d C = - i\; {\cal B},\qquad\qquad s_d{\cal B} = 0,\nonumber\\
&& s_d E = D_\mu\partial^\mu\bar C,\qquad s_d({\partial_\mu A^\mu})= 0,\quad\quad
s_d \,{\bar \lambda} = -i\;({\partial_\mu A^\mu}),\quad\quad s_d\,{\lambda } = 0.
\end{eqnarray}
 It can be checked that the above (anti-)co-BRST symmetry transformations are 
 off-shell nilpotent and absolutely anticommuting in nature (where we do not
 invoke any kinds of CF-type restrictions for its validity). We {\it also} note that
 the superscripts ($\lambda$) and ($\bar\lambda$) on the Lagrangian densities
 are logically correct because the Lagrange multipliers $\lambda$ and $\bar\lambda$
 characterize these Lagrangian densities. Furthermore, we observe that these Lagrange multiplier
 fields carry the ghost numbers equal to (+1) and (-1), respectively. Finally, it can be explicitly
 checked that the following are true, namely;
 \begin{eqnarray}
&&s_d {\cal L}_B^{(\bar\lambda)} = \partial_{\mu}[{\cal B}\cdot\partial^{\mu}\bar C],
\qquad\qquad s_{ad} {\cal L}^{(\lambda)}_{\bar B} = 
\partial_{\mu}[{\cal B}\cdot\partial^{\mu} C],\nonumber\\
&&s_d {\cal L}^{(\lambda)}_{\bar B} = \partial_{\mu}[{\cal B}\cdot D^{\mu}\bar C
-\varepsilon^{\mu\nu}(\partial_\nu\bar C\times\bar C)\cdot C ],\nonumber\\
&& s_{ad}{\cal L}_B^{(\bar\lambda)} =\partial_{\mu}[{\cal B}\cdot D^{\mu}  C
+ \varepsilon^{\mu\nu} \bar C\cdot(\partial_{\nu} C\times C)],
\end{eqnarray}
 which demonstrate that the action integrals $ S = \int d^2 x\; {\cal L}_B^{(\bar\lambda)}$ and $ S = \int d^2x\; {\cal L}_{\bar B}^{(\lambda)}$
 remain invariant under the (anti-)co-BRST symmetry transformations. We would like to lay emphasis on the fact that
 {\it both} the Lagrangian densities ${\cal L}_B^{(\bar\lambda)}$ and ${\cal L}_{\bar B}^{(\lambda)}$
 respect {\it both }  the co-BRST and anti-co-BRST symmetries (cf. Eq. (40)), separately and independently.

 A close look at the transformations (41) and (8) (cf. Sec. 2) demonstrates that the expressions for the charge
 $Q^{(\bar\lambda)}_d = Q_d$ and $Q_{ad}^{(\lambda)} = Q_{ad}$ (cf. Eq. (8)) remain the {\it same}
 as far as the Lagrangian densities in (1) {\it and} ${\cal L}_B^{(\bar\lambda)}$ as well as ${\cal L}_{\bar B}^{(\lambda)}$
are concerned. However, we note that the anti-co-BRST charge $Q_{ad}^{(\bar\lambda)}$(derived from the Lagrangian
density ${\cal L}_B^{(\bar\lambda)}$) and co-BRST charge $Q_d^{(\lambda)}$  (derived from the Lagrangian density 
${\cal L}_{\bar B}^{(\lambda)}$) would be different from (8). These conserved charges
and their expressions have been derived in our earlier work (see, e.g. [16] for details). 
We quote here these expressions explicitly:
\begin{eqnarray}
Q^{(\bar\lambda)}_{ad} &= &\int\; dx\; \Big[ {\cal B}\cdot\dot C - \partial_1 B\cdot C
+ \bar C\cdot (\partial_1 C\times C)\Big]\nonumber\\
&\equiv & \int\; dx\,\Big[{\cal B}\cdot\dot C - D_0{\cal B}\cdot C + (\partial_1 \bar C\times C)\cdot C
+ \bar C\cdot (\partial_1 C\times C)\Big]\nonumber\\
&\equiv & \int\; dx\,\Big[{\cal B}\cdot\dot C - D_0{\cal B}\cdot C 
- \bar C\cdot (\partial_1 C\times C)\Big],\nonumber\\
Q_d^{(\lambda)} & = & \int\;dx\;\Big[{\cal B}\cdot\dot{\bar C} + \partial_1 \bar B\cdot\bar C - (\partial_1\bar C\times\bar C)\cdot C\Big]\nonumber\\
&\equiv & \int \; dx\;\Big[ {\cal\ B}\cdot\dot{\bar C} - D_0{\cal B}\cdot\bar C
- (\bar C\times\partial_1 C)\cdot\bar C - (\partial_1\bar C\times\bar C)\cdot C\Big]\nonumber\\
&\equiv & \int \; dx\;\Big[ {\cal\ B}\cdot\dot{\bar C} - D_0{\cal B}\cdot\bar C
+ (\bar C\times\partial_1 \bar C)\cdot C \Big].  
\end{eqnarray}
In the above equivalent expressions, we have utilized the equations of motion (derived from the Lagrangian  densities
in Eq. (39)) and we have {\it also} dropped the total space derivative terms.  
To prove the nilpotency  [$(Q_{ad}^{(\bar\lambda)})^ 2 = 0,  (Q_d^{(\lambda)})^ 2 = 0 $]
of the above charges, we note that they can be expressed in terms of the (anti-)co-BRST transformations as:
\begin{eqnarray}
&&Q_{ad}^{(\bar\lambda)} = s_{ad}\;\Big(\int\; dx\;\big [ -\;i\;\bar C\cdot D_0 C +  i\; \dot{\bar C}\cdot C\big]\Big),\nonumber\\
&&Q_d^{(\lambda)} = s_d\; \Big( \int\; dx\; \big[ i\; \bar C\cdot \dot C - i\;D_0\bar C\cdot C\big]\Big).
\end{eqnarray}
The above expressions for the (anti-)co-BRST charges produce the last
entry in the  expressions for the charges $Q_{ad}^{(\bar\lambda)}$
and $ Q_d^{(\lambda)}$ in Eq. (42). It can be now trivially checked that:
\begin{eqnarray}
&& s_{ad}\; Q_{ad}^{(\bar\lambda)} = -\;i\;{\{ Q_{ad}^{(\bar\lambda)}, Q_{(ad)}^{(\bar\lambda)}}\} = 0\qquad\Longleftrightarrow\qquad s_{ad}^ 2 = 0,\nonumber\\
&& s_d \;Q_d^{(\lambda)} ~= -\;i\;{\{Q_d^{(\lambda)}, Q_d^{(\lambda)}}\} = 0~\qquad\Longleftrightarrow\qquad s_d^2 ~= 0.
\end{eqnarray}
Thus, we observe that the nilpotency of the charges $Q_{ad}^{(\bar\lambda)}$ and $Q_d^{(\lambda)}$
is deeply connected with the nilpotency of the (anti-)-co-BRST symmetries (i.e. $s_{(a)d}^ 2 = 0$)
when we exploit the beauty and strength  of the connection between the continuous symmetries  and their corresponding 
generators. We would like to state that the nilpotency of the charges $ Q_d^{(\bar\lambda)}  = Q_d $ (cf. Eqs. (8) and (9)) and 
$Q_{ad}^{(\lambda)} = Q_{ad}$ have already been  proven in Eq. (9). This happens because of the fact
that the expressions for $Q_{ad}^{(\lambda)}$ and $Q_d^{(\bar\lambda)}$ are same as 
given in Eq. (8) for the Lagrangian densities (1). Thus, we have proven the nilpotency of {\it all} the
charges derived from the modified Lagrangian densities (39) where  $\lambda$ and $\bar\lambda$ are present.

We now focus on the proof of the property of absolute anticommutativity
of the charges  $Q_{ad}^{(\bar\lambda)}$ and $Q_d^{(\lambda)}$ which are non-trivial (cf. Eq. (42)). In this connection,
we would like to point out that the absolute anticommutativity 
of the charges $ Q_d^{(\bar\lambda)}  = Q_d $ and $Q_{ad}^{(\lambda)} = Q_{ad}$ 
has already been proven in our present section itself.
We note that the following are true:
\begin{eqnarray}
&&Q_d^{(\lambda)} = s_{ad}\Big[\int\; dx\; \big(- i\;\bar C\cdot\dot{\bar C} + \frac {\bar C}{2}\cdot (A_0\times\bar C)\big)\Big],\nonumber\\
&&Q_{ad}^{(\bar\lambda)}= s_d\Big[\int\; dx \;\big(i\; C\cdot\dot C - \frac { C}{2}\cdot (A_0\times C)\big)\Big].
\end{eqnarray}
The above expressions demonstrate that the absolute anticommutativity property
of the (anti-)co-BRST charges (i.e. ${\{Q_d^{(\lambda)}, Q_{ad}^{(\bar\lambda)}}\} = 0$)
is {\it true} and this property is primarily connected with the off-shell nilpotency  $(s_{(a)d}^2 = 0)$
of the (anti-)co-BRST symmetry transformations $(s_{(a)d})$ that are present in our 2D non-Abelian theory (cf. Eq. (40)).
To corroborate the above statements, it is straightforward to note that:
\begin{eqnarray}
&& s_{ad}\; Q_d^{(\lambda)} = - i\; {\{Q_d^{(\lambda)}, Q_{ad}^{(\bar\lambda)}}\} = 0\quad\Longleftrightarrow\quad s_{ad}^2 = 0,\nonumber\\
&& s_d\; Q_{ad}^{(\bar\lambda)} = - i \;{\{Q_{ad}^{(\bar\lambda)}, Q_d^{(\lambda)}}\} = 0~\quad\Longleftrightarrow \quad s_d^2 = 0.
\end{eqnarray}
From the above relationships, it is crystal clear that the absolute 
anticommutativity (i.e. ${\{Q_d^{(\lambda)}, Q_{ad}^{(\bar\lambda)}}\} = 0$) for the 
(anti-)co-BRST charges is deeply connected with the nilpotency $(s_{(a)d}^2 = 0)$
property of the (anti-)co-BRST symmetry transformations $(s_{(a)d})$ for the Lagrangian densities 
(39). We wrap up this section with the remark that we have proven the nilpotency and absolute 
anticommutativity properties of the (anti-)co-BRST charges for the Lagrangian densities (1)
as well as (39) where we do {\it not} invoke any kinds of CF-type restrictions. This observation 
 is {\it novel} and drastically different from the proof of the absolute anticommutativity property 
 of the  conserved and nilpotent (anti-)BRST charges where it is mandatory for us to invoke the CF-condition.

,

\section{Nilpotency and Absolute Anticommutativity of the Fermionic Charges: Superfield Approach}

We express here the properties of nilpotency and absolute anticommutativity by exploiting the geometrical 
AVSA to BRST formalism. In this connection, first of all, we recall 
that the (anti-)BRST symmetry transformations $s_{(a)b}$
have been shown to be connected with the translational generators  
$(\partial_{\theta}, \partial_{\bar\theta})$  
along $(\theta,\bar\theta)$-directions of the (2, 2)-dimensional supermanifold through the 
following mappings:
\begin{eqnarray}
&& s_b\longleftrightarrow \frac{\partial}{\partial\bar\theta}\Big|_{\theta = 0},\qquad s_{ab}\longleftrightarrow \frac{\partial}{\partial\theta}\Big|_{\bar\theta = 0}.
\end{eqnarray}
We can very well choose the Grassmannian  variables to be $(\theta_1, \theta_2)$ and
identify the nilpotent symmetries: $s_b \leftrightarrow \partial_{\theta_1}|_{\theta_2 = 0}$ and $s_{ab} \leftrightarrow \partial_{\theta_2}|_{\theta_1 = 0}$
because there are other nilpotent $(s_{(a)d}^2 = 0)$ symmetries $s_{(a)d}$ in our theory, too. The {\it latter}
nilpotent symmetries could be identified with translational generators as: $s_d \leftrightarrow \partial_{\theta_3}|_{\theta_4 = 0}$ and 
$s_{ad} \leftrightarrow \partial_{\theta_4}|_{\theta_3 = 0}$ where we shall have another set of a pair of Grassmannian
variables $(\theta_3, \theta_4)$. However, for the sake of brevity, we have chosen only  $(\theta, \bar \theta)$ as the
Grassmannian variables so that we could discuss the (anti-)BRST and (anti-)co-BRST symmetries, separately and independently.
The above mappings imply that the nilpotency of the (anti-)BRST symmetries  (i.e. $s_{(a)b}^{2} = 0$) is intimately connected with 
the nilpotency ($\partial_{\theta}^{2} = \partial_{\bar\theta}^{2} = 0$)  of the translational generators
 ($\partial_{\theta} , \partial_{\bar\theta}$).   
This observation is utilized in expressing the expressions for the conserved and nilpotent
(anti-)BRST charges in (33) as follows:
\begin{eqnarray}
Q_{ab}  & = & \frac{\partial}{\partial\theta}\;\int\;dx\;\Big [i\, \bar F^{(h)}(x,\theta,\bar\theta)\cdot
\dot F^{(h)}(x,\theta,\bar\theta) - \bar B(x)\cdot B_0^{(h)}(x,\theta,\bar\theta)\Big]\Big|_{\bar\theta = 0}\nonumber\\
& \equiv & \int d\theta\;\int dx\; \Big [i\,\bar F^{(h)}(x,\theta,\bar\theta)\cdot
\dot F^{(h)}(x,\theta,\bar\theta) - \bar B(x)\cdot B_0^{(h)}(x,\theta,\bar\theta)\Big]\Big|_{\bar\theta = 0},\nonumber\\
Q_b & = &\frac{\partial}{\partial\bar\theta}\int dx\;\Big[ B(x)\cdot B_0^{(h)}(x,\theta,\bar\theta)
  + i\;\dot{\bar F}^{(h)}(x,\theta,\bar\theta)\cdot F^{(h)}(x,\theta,\bar\theta)\Big]\Big|_{\theta = 0}\nonumber\\
&\equiv & \int\;d\bar\theta\;\int dx\;\Big [B(x)\cdot B_0^{(h)}(x,\theta,\bar\theta)
+ i\;\dot{\bar F}^{(h)}(x,\theta,\bar\theta)\cdot F^{(h)}(x,\theta,\bar\theta)\Big ]\Big|_{\theta = 0}.
\end{eqnarray}
The above expressions establish the nilpotency of the (anti-)BRST charges $ Q_{(a)b}$ because:
\begin{eqnarray}
&&\partial_{\theta} Q_{ab} = 0~~\Longleftrightarrow ~~~\partial_{\theta}^ 2 
= 0~~~\Longleftrightarrow ~~~ s_{ab} Q_{ab} = - i\; {\{Q_{ab}, Q_{ab}}\} = 0,\nonumber\\
&&\partial_{\bar\theta} Q_b = 0~~~\Longleftrightarrow ~~~\partial_{\bar\theta}^ 2 = 0~~~ 
\Longleftrightarrow~~~ s_b Q_b = - i\; {\{Q_b, Q_b}\} = 0.
\end{eqnarray}
It should be noted that we do {\it not} invoke any kinds 
of CF-type restrictions for the proof of off-shell nilpotency of the above (anti-)BRST
charges.

We capture now the absolute anticommutativity property 
of the (anti-)BRST symmetry generators $Q_{(a)b}$ in the language 
of the AVSA to BRST formalism. In this context, 
we concentrate on the expressions for (anti-)BRST charges that have been
quoted in (35). It can be checked that we have the following expressions for these charges in the language 
of the AVSA to BRST formalism, namely;
\begin{eqnarray}
Q_{ab} &=& \frac{\partial}{\partial \bar\theta}\; \Big[\int dx\; \Big\{ -\;i\;{\bar F}^{(h)}(x,\theta,\bar\theta)\cdot \dot{\bar F}^{(h)}(x,\theta,\bar\theta)\nonumber\\ 
&+& \frac {1}{2}\;{\bar F}^{(h)}(x,\theta,\bar\theta)\cdot(B_0^{(h)}(x,\theta,\bar\theta)
\times {\bar F}^{(h)}(x,\theta, \bar\theta))\Big\}\Big]\Big|_{\theta = 0}
\nonumber\\
& = &\int\;d\bar\theta\;\Big[\int dx\; \Big\{ -\;i\;{\bar F}^{(h)}(x,\theta,\bar\theta)\cdot \dot{\bar F}^{(h)}(x,\theta,\bar\theta)\nonumber\\ 
&+& \frac {1}{2}\;{\bar F}^{(h)}(x,\theta,\bar\theta)\cdot(B_0^{(h)}(x,\theta,\bar\theta)
\times {\bar F}^{(h)}(x,\theta, \bar\theta))\Big\}\Big]\Big|_{\theta = 0},
\nonumber\\
Q_b &=& \frac{\partial}{\partial \theta}\;\Big[\int dx\; \Big\{ \;i\;F^{(h)}(x,\theta,\bar\theta)\cdot \dot F^{(h)}(x,\theta,\bar\theta)\nonumber\\ 
&-& \frac {1}{2}\; F^{(h)}(x,\theta,\bar\theta)\cdot(B_0^{(h)}(x,\theta,\bar\theta)
\times  F^{(h)}(x,\theta, \bar\theta))\Big\}\Big]\Big|_{\bar\theta = 0}
\nonumber\\
& = & \int d\theta\; \Big[\int dx\; \Big\{ \;i\;F^{(h)}(x,\theta,\bar\theta)\cdot \dot F^{(h)}(x,\theta,\bar\theta)\nonumber\\ 
&-& \frac {1}{2}\; F^{(h)}(x,\theta,\bar\theta)\cdot(B_0^{(h)}(x,\theta,\bar\theta)
\times  F^{(h)}(x,\theta, \bar\theta))\Big\}\Big]\Big|_{\bar\theta = 0}.
\end{eqnarray}
It is straightforward  to note that the nilpotency properties of the 
translational generators $(\partial_\theta, \partial_{\bar\theta})$ 
along the Grassmannian directions imply that:
\begin{eqnarray}
&&\partial_{\bar\theta} Q_{ab} = 0~~~\Longleftrightarrow~~~\partial_{\bar\theta}^2 = ~0,\qquad\quad\partial_\theta Q_b = 0 ~~~
\Longleftrightarrow
~~~ \partial_\theta^2 = 0.  
\end{eqnarray} 
The above observations lead us to draw the conclusion that the absolute anticommutativity $(Q_b Q_{ab} + Q_{ab} Q_b = 0)$
of the (anti-)BRST charges (cf. Eq. (36)) in the ordinary space can be captured in the language of the 
superfield approach to BRST formalism.

We briefly comment here on the expressions for the (anti-)co-BRST charges 
$Q_{(a)d}$ that have been expressed in {\it two} different ways in Eq. (33) and Eq. (35).
We have established earlier that the following mappings 
are true in the  cases of $s_d$ and $s_{ad}$:
\begin{eqnarray}
&& s_d \longleftrightarrow  \lim _{\theta~ = ~ 0} \frac {\partial}{\partial\bar\theta},\qquad\qquad s_{ad}\longleftrightarrow \lim _{\bar\theta~ =~0}\frac {\partial}{\partial\theta}.
\end{eqnarray}
We can very 
well repeat here the previous footnote written in our manuscript. However, this would be 
{\it only} an academic exercise. The main issue is the fact that we discuss the (anti-)BRST
and (anti-)co-BRST symmetries, within the framework of AVSA to BRST formalism,
separately and independently. Thus, when we focus on $(\theta_1, \theta_2)$, 
we do not bother about $(\theta_3, \theta_4)$ and {\it vice-versa}. This is precisely the reason
that we have taken, for the sake of brevity,  {\it only} the (2, 2)-dimensional supermanifold for our discussion where,
at a time, only a pair of Grassmannian variables are taken into account.
Thus, the nilpotency of the (anti-)co-BRST charges can be expressed in terms of the 
quantities on the (2, 2)-dimensional supermanifold as follows:
\begin{eqnarray}
Q_d  &  = & \frac{\partial}{\partial\bar\theta}\;\Big[\int dx\; \Big {\{{\cal B}(x)\cdot  B_1^{(dg)}(x,\theta,\bar\theta)
+ B(x)\cdot B_0^{(dg)} (x, \theta,\bar\theta)}\Big\}\Big]\Big|_{\theta =~0}\nonumber\\
&\equiv & \int d\bar\theta\int dx \;\Big[ {\cal B}(x)\cdot  B_1^{(dg)}(x,\theta,\bar\theta)
+ B(x)\cdot B_0^{(dg)} (x, \theta,\bar\theta)\Big]\Big|_{\theta =~0},
\end{eqnarray}
where the superscript $(dg)$ denotes the superfields (cf. Eq. (32)) that have been obtained after the application of 
(anti-)co-BRST invariant restriction in Eq. (29). Similarly, we note that the following is correct, namely;
\begin{eqnarray}
Q_{ad} & = &\frac{\partial}{\partial\theta}\;\Big [\int dx\; \Big {\{{\cal B}(x)\cdot B_1^{(dg)}(x,\theta,\bar\theta)
- \bar B(x)\cdot B_0^{(dg)}(x,\theta,\bar\theta)}\Big\}\Big]\Big|_{\bar\theta =~0}\nonumber\\
& \equiv & \int d\theta \int dx\; \Big[{\cal B}(x)\cdot B_1^{(dg)}(x,\theta,\bar\theta)
- \bar B(x)\cdot B_0^{(dg)}(x,\theta,\bar\theta)\Big]\Big|_{\bar\theta =~0}.
\end{eqnarray}
It is crystal clear, from Eqs. (53) and (54), that the following are true:
\begin{eqnarray}
&&\partial_{\bar\theta}\; Q_d  = 0\quad\Longleftrightarrow\quad\partial_{\bar\theta}^2 = 0,\qquad \partial_{\theta}\; Q_{ad} = 0\quad\Longleftrightarrow \quad \partial_{\theta}^2 = 0. 
\end{eqnarray}
The above relationships, in the ordinary 2D space, correspond to the following explicit expressions
in the language of anticommutators:
\begin{eqnarray}
&&s_d\; Q_d =  -\;i\; {\{Q_d, Q_d}\} = 0~~~~~~\Longleftrightarrow ~~ Q_d^2 = 0\qquad\Longleftrightarrow \quad s_d^2 = 0,\nonumber\\
&& s_{ad}\;Q_{ad} = -\;i\;{\{Q_{ad}, Q_{ad}}\} = 0~~\Longleftrightarrow~~  Q_{ad}^2 = 0~\quad\Longleftrightarrow\quad s_{ad}^2 = 0.
\end{eqnarray}
Thus, we have captured the nilpotency property of the (anti-)co-BRST charges in the language 
of the quantities that are defined on the (2, 2)-dimensional supermanifold.
In fact, the nilpotency $(Q_{(a)d}^2  = 0)$ of the (anti-)co-BRST charges is
deeply connected with the nilpotency $(\partial_{\theta}^2 = 0, \partial_{\bar\theta}^2 = 0)$
of the translational generators $(\partial_{\theta},\partial_{\bar\theta})$
along the Grassmannian directions $(\theta, \bar\theta)$ of the (2, 2)-dimensional supermanifold.

Now we dwell a bit on the absolute anticommutativity property
of the (anti-)co-BRST charges $Q_{(a)d}$ that have been expressed in Eq. (35).
Taking the inputs from Eqs. (52), (32) and (26), we have the following 
\begin{eqnarray}
 Q_{ad} & = & \frac {\partial}{\partial\bar\theta}\;\Big[\int\; dx\; \big (i \; F^{(dh)}(x,\theta,\bar\theta)\cdot\dot F^{(dh)}(x,\theta,\bar\theta)\big)\nonumber\\
  & - & \frac {1}{2}\;F^{(dh)}(x,\theta,\bar\theta)\cdot \big(B_0^{(dg)}(x,\theta,\bar\theta)\times F^{(dh)}(x,\theta,\bar\theta)\big)\Big]\Big|_{
\theta = 0}\nonumber\\
& \equiv &  \int d\bar\theta\; \Big[\int\; dx\; \big (i \; F^{(dh)}(x,\theta,\bar\theta)\cdot\dot F^{(dh)}(x,\theta,\bar\theta)\big)\nonumber\\
  & - & \frac {1}{2}\;F^{(dh)}(x,\theta,\bar\theta)\cdot \big(B_0^{(dg)}(x,\theta,\bar\theta)\times F^{(dh)}(x,\theta,\bar\theta)\big)\Big]\Big|_{
\theta = 0}\nonumber\\
 Q_d & = & \frac {\partial}{\partial\theta}\;\Big[\int\; dx\; \big (- i\; \bar F^{(dh)}(x,\theta,\bar\theta)\cdot\dot{\bar F}^{(dh)}(x,\theta,\bar\theta)\big)\nonumber\\
 & + & \frac {1}{2}\;\bar F^{(dh)}(x,\theta,\bar\theta)\cdot \big(B_0^{(dg)}(x,\theta,\bar\theta)\times\bar F^{(dh)}(x,\theta,\bar\theta)\big)\Big]\Big|_{\bar\theta = 0}\nonumber\\
  & \equiv & \int d\theta\;\int dx\;\Big[\int\; dx\; \big (- i\; \bar F^{(dh)}(x,\theta,\bar\theta)\cdot\dot{\bar F}^{(dh)}(x,\theta,\bar\theta)\big)\nonumber\\
 & + & \frac {1}{2}\;\bar F^{(dh)}(x,\theta,\bar\theta)\cdot \big(B_0^{(dg)}(x,\theta,\bar\theta)\times\bar F^{(dh)}(x,\theta,\bar\theta)\big)\Big]\Big|_{\bar\theta = 0}.
\end{eqnarray}
Thus, the expressions for the (anti-)co-BRST charges (cf. Eq. (57)) imply
\begin{eqnarray}
&&\partial_{\theta}\; Q_d = 0\quad\Longleftrightarrow \quad \partial_{\theta}^2 = 0\quad\Longleftrightarrow \quad s_{ad}\; Q_d =  - i\; {\{Q_d, Q_{ad}}\} = 0,\nonumber\\
&& \partial_{\bar\theta}\; Q_{ad} = 0\quad\Longleftrightarrow \quad \partial_{\bar\theta}^2 = 0\quad\Longleftrightarrow \quad s_d\;Q_{ad} = - i\; {\{Q_{ad},Q_d}\} = 0.
\end{eqnarray}
The above expressions capture the absolute anticommutativity property of the (anti-)co-BRST charge (i.e.  ${\{Q_d, Q_{ad}}\} = 0)$
in the language of AVSA to BRST formalism. We observe, once again, it is the nilpotency $(\partial_{\theta}^2 =\partial_{\bar\theta}^2 = 0)$
of the translational generators $(\partial_{\theta}, \partial_{\bar\theta})$ that plays a decisive role
in capturing the nilpotency as well as absolute anticommutativity properties of the (anti-)co-BRST charges 
in the terminology of AVSA to BRST formalism.

Finally, we would like to comment briefly on the nilpotency and absolute anticommutativity
properties of the (anti-)co-BRST charges $(Q_d ^{(\lambda)}, Q_{ad}^{(\bar\lambda)})$
that have been derived from the Lagrangian densities (39) and listed in (42)
in different forms. We would  like to lay emphasis on the fact that the Lagrangian densities (39) are 
very {\it special} in the sense that these Lagrangian densities respect proper (anti-)co-BRST symmetry transformations
(listed in Eq. (40)) separately and independently (cf. Eq. (41)) where we do {\it not} invoke any kinds
of CF-type restrictions from outside. Thus, as far as symmetry considerations are concerned, these Lagrangian
densities are really beautiful from the point of view of the proper (anti-)co-BRST symmetry transformations (41).   
Within the framework of AVSA to BRST formalism, it can be 
checked that the expressions in (43) are:
\begin{eqnarray}
Q_{ad}^{(\bar\lambda)} & = &\frac {\partial}{\partial\theta}\;\int dx \;\Big[ - i\;\bar F^{(dh)}(x,\theta,\bar\theta)\cdot\dot F^{(dh)}(x,\theta,\bar\theta)\nonumber\\
& + &\bar F^{(dh)}(x,\theta,\bar\theta)\cdot (B_0^{(dg)}(x,\theta,\bar\theta)\times F^{(dh)}(x,\theta,\bar\theta))\nonumber\\
& + & i\;\bar F^{(dh)}(x,\theta,\bar\theta)\cdot F^{(dh)}(x,\theta,\bar\theta)\Big]\Big|_{\bar\theta~ =~ 0}\nonumber\\
&\equiv & \int d\theta\;\int dx\; \Big[ - i\;\bar F^{(dh)}(x,\theta,\bar\theta)\cdot\dot F^{(dh)}(x,\theta,\bar\theta)\nonumber\\
& + &\bar F^{(dh)}(x,\theta,\bar\theta)\cdot (B_0^{(dg)}(x,\theta,\bar\theta)\times F^{(dh)}(x,\theta,\bar\theta))\nonumber\\
& + & i\;\bar F^{(dh)}(x,\theta,\bar\theta)\cdot F^{(dh)}(x,\theta,\bar\theta)\Big]\Big|_{\bar\theta~ =~ 0},\nonumber\\
Q_{d}^{(\lambda)} & = &\frac {\partial}{\partial\bar\theta}\;\int dx \;\Big[  i\;\bar F^{(dh)}(x,\theta,\bar\theta)\cdot\dot F^{(dh)}(x,\theta,\bar\theta),\nonumber\\
& - &\dot {\bar F}^{(dh)}(x,\theta,\bar\theta)\cdot  F^{(dh)}(x,\theta,\bar\theta)\nonumber\\
& + & (B_0^{(dg)}(x,\theta,\bar\theta)\times \bar F^{(dh)}(x,\theta,\bar\theta))\cdot F^{(dh)}(x,\theta,\bar\theta)\Big]\Big|_{\theta~ =~ 0}\nonumber\\
&\equiv & \int d\bar\theta\;\int dx\; \Big[ i\;\bar F^{(dh)}(x,\theta,\bar\theta)\cdot\dot F^{(dh)}(x,\theta,\bar\theta)\nonumber\\
& - &\dot {\bar F}^{(dh)}(x,\theta,\bar\theta)\cdot  F^{(dh)}(x,\theta,\bar\theta)\nonumber\\
& + & (B_0^{(dg)}(x,\theta,\bar\theta)\times \bar F^{(dh)}(x,\theta,\bar\theta))\cdot F^{(dh)}(x,\theta,\bar\theta)\Big]\Big|_{\theta~ =~ 0},
\end{eqnarray}
where the superfields with superscripts $(dh)$ and $(dg)$ have been
explained in Sec. 4.
It is clear, from the above expressions, that we have the following:
\begin{eqnarray}
 \partial_{\theta}\; Q_{ad}^{(\bar\lambda)} &=& 0\qquad \Longleftrightarrow \nonumber\\
\partial_{\theta}^2 &=& 0\qquad\Longleftrightarrow \nonumber\\
s_{ad}\; Q_{ad}^{(\bar\lambda)} &=& - \;i {\{ Q_{ad}^{(\bar\lambda)}, Q_{ad}^{(\bar\lambda)}}\} = 0,\nonumber\\
\partial_{\bar\theta}\; Q_d^{(\lambda)} &=& 0\qquad \Longleftrightarrow \nonumber\\
 \partial_{\bar\theta}^2 &=& 0\qquad\Longleftrightarrow \nonumber\\
s_d\; Q_d^{(\bar\lambda)} &=& - \;i {\{ Q_d^{(\lambda)}, Q_d^{(\lambda)}}\} = 0.
\end{eqnarray}
The above relations prove the nilpotency of $Q_d^{(\lambda)}$ and $Q_{ad}^{(\bar\lambda)}$ which is also connected with the
nilpotency of the translational generators $(\partial_\theta, \partial_{\bar\theta})$  along the Grassmannian directions 
$(\theta, \bar\theta)$ of the (2, 2)-dimensional supermanifold on which our 2D {\it ordinary} theory is considered.
To be more precise, the nilpotency of the above (anti-)co-BRST charges (which have been derived from the Lagrangian densities (39))
becomes very transparent when concentrating on the {\it third} and {\it sixth} lines in  (60).
In particular, the anticommutator of the conserved charges with {\it themselves} being zero
 immediately implies the nilpotency property (of these conserved charges). 
Let us now concentrate on the forms of the (anti-)co-BRST charges that have been written in Eq. (45).
As is evident from Eq. (46), the absolute anticommutativity property of the 
(anti-)co-BRST charges is primarily hidden in Eq. (45) and is 
deeply connected with the nilpotency property of the (anti-)co-BRST symmetry transformations $(s_{(a)d})$.
Thus, we express the forms of the (anti-)co-BRST charges (45) in the 
language of AVSA to BRST formalism as follows
\begin{eqnarray}
Q_{ad}^{(\bar\lambda)} & = &\frac {\partial}{\partial\bar\theta}\;\int dx\; \Big [  i\;  F^{(dh)}(x,\theta,\bar\theta)\cdot\dot F^{(dh)}(x,\theta,\bar\theta)\nonumber\\
& - & \frac {1}{2}\;  F^{(dh)}(x,\theta,\bar\theta)\cdot (B_0^{(dg)}(x,\theta,\bar\theta)\times  F^{(dh)}(x,\theta,\bar\theta))\Big]\Big|_{\theta ~=~ 0}\nonumber\\
& \equiv & \int d\bar\theta\;\int dx\;\Big [  i\;  F^{(dh)}(x,\theta,\bar\theta)\cdot\dot F^{(dh)}(x,\theta,\bar\theta)\nonumber\\
& - & \frac {1}{2}\;  F^{(dh)}(x,\theta,\bar\theta)\cdot (B_0^{(dg)}(x,\theta,\bar\theta)\times  F^{(dh)}(x,\theta,\bar\theta))\Big]\Big|_{\theta ~=~ 0},
\nonumber\\
Q_d^{(\lambda)} & = &\frac {\partial}{\partial\theta}\;\int dx\; \Big [ - i\; \bar F^{(dh)}(x,\theta,\bar\theta)\cdot\dot{\bar F}^{(dh)}(x,\theta,\bar\theta)\nonumber\\
& + & \frac {1}{2}\; \bar F^{(dh)}(x,\theta,\bar\theta)\cdot (B_0^{(dg)}(x,\theta,\bar\theta)\times \bar F^{(dh)}(x,\theta,\bar\theta))\Big]\Big|_{\bar\theta~ = ~0}\nonumber\\
& \equiv & \int d\theta\;\int dx\;\Big [ - i\; \bar F^{(dh)}(x,\theta,\bar\theta)\cdot\dot{\bar F}^{(dh)}(x,\theta,\bar\theta)\nonumber\\
& + & \frac {1}{2}\; \bar F^{(dh)}(x,\theta,\bar\theta)\cdot (B_0^{(dg)}(x,\theta,\bar\theta)\times \bar F^{(dh)}(x,\theta,\bar\theta))\Big]\Big|_{\bar\theta ~=~ 0},
\end{eqnarray}
where the superfields with superscripts $(dh)$ and $(dg)$ have already been explained in Sec. 4. It is straightforward
to note, from the above equation, that:
\begin{eqnarray}
&&\partial_{\theta}\; Q_d^{(\lambda)} = 0\quad\Longleftrightarrow \quad \partial_{\theta}^2 = 0\quad\Longleftrightarrow \quad s_{ad}\; Q_d^{(\lambda)}
 = - i\;{\{Q_d^{(\lambda)},Q_{ad}^{(\bar\lambda)}}\} = 0,\nonumber\\
&&\partial_{\bar\theta}\; Q_{ad}^{(\bar\lambda)} =  0\quad\Longleftrightarrow \quad \partial_{\bar\theta}^2 = 0\quad\Longleftrightarrow \quad s_d\;Q_{ad}^{(\bar\lambda)} = - i\; {\{Q_{ad}^{(\bar\lambda)}, Q_d^{(\lambda)}}\} = 0.
\end{eqnarray}
We end this section with the remark that the absolute anticommutativity property of the (anti-)co-BRST charges is deeply connected 
with the nilpotency property $ (\partial_{\theta}^2 = \partial_{\bar\theta}^2 = 0)$ of the translational generators $ (\partial_{\theta}, \partial_{\bar\theta})$ along the Grassmannian directions $(\theta,\bar\theta)$ of the (2, 2)-dimensional supermanifold on
which our 2D ordinary non-Abelian theory is generalized.

\section{Conclusions}

\noindent
We have exploited the theoretical  strength of the AVSA to BRST
formalism to express the properties of the nilpotency and absolute anticommutativity
of the fermionic conserved charges (i.e. (anti-)BRST and (anti-)co-BRST charges)
of our self-interacting 2D non-Abelian theory (without any interaction  with matter fields).
We have {\it not} achieved this goal in our earlier works [9-12] on the AVSA to BRST formalism.
Thus, the results in our present investigation are achieved for the first time. It is straightforward to express
the nilpotency property of the fermionic (i.e. conserved (anti-)BRST and (anti-)co-BRST) charges in the language 
of the AVSA to BRST formalism. However, the property
 of absolute anticommutativity is captured, within the framework
 of AVSA to BRST
formalism, by applying specific mathematical 
trick where the CF-condition plays a decisive role.

We would like to lay emphasis on the contents of Sec. 5 where we have been able to exploit the 
virtues of  symmetry principles to express the (anti-)BRST and (anti-)co-BRST charges in various
{\it exact-forms}. These theoretical  expressions have been exploited, in turn, to capture
the nilpotency and absolute anticommutativity properties in the language of AVSA to BRST 
formalism in Sec. 6. We observe that the CF-condition $(B + \bar B +  (C\times \bar C) = 0)$
enables us in expressing the BRST charge as an anti-BRST {\it exact} form and anti-BRST charge as a 
BRST {\it exact} form. We would like to lay emphasis on the fact that the contents of Secs. 5 and 6
are intertwined in an elegant manner. Though it appears, from our statements in this paragraph, that
the contents of Sec. 5 have influenced our results in Sec. 6. However, we would like to stress that,
many a times, our understandings of the contents of Sec. 6 have influenced our results in Sec. 5. Thus,
to be precise, the key results of our present endeavor are influenced by our knowledge of {\it both} the above sections
which are inter-related. 
These results play an important role in establishing
the absolute anticommutativity properties of the above fermionic charges. Thus, first of all, we have proven the nilpotency 
and absolute anticommutativity of  the (anti-)BRST and (anti-)co-BRST charges in the language 
of symmetry properties (cf. Sec. 5). In particular, we have 
shown that it is the nilpotency  of the (anti-)BRST
and (anti-)co-BRST symmetry transformations that has played a decisive role in the 
proof of the above {\it properties} in the {\it ordinary } 2D space of our
non-Abelian 1-form gauge theory. In fact, the results of Sec. 5 have been translated into the
language of AVSA to BRST formalism in Sec. 6.

The proof of the nilpotency and absolute anticommutativity properties in the language of 
AVSA to BRST formalism for the fermionic (anti-)BRST and (anti-)co-BRST charges 
is a {\it novel} result because, in our earlier  works on AVSA to BRST formalism [9-12],
we have not achieved this goal. In our very recent works [30-32], we have captured 
the property of the absolute anticommutativity of nilpotent charges within the framework of 
(anti-)chiral superfield approach to BRST formalism. However, we have {\it not}
done {\it so} within the framework of AVSA to BRST formalism where the {\it full} expansions 
of the superfields are taken into account. We plan to exploit our present idea
to consolidate it by applying it to the cases of 1D toy model of the rigid rotor,
2D self-dual bosonic theory, modified versions of 2D Proca and anomalous gauge 
theory, 4D Abelian 2-form and 6D Abelian 3-form gauge theories where we have demonstrated
the existence of (anti-)BRST and (anti-)co-BRST charges as these theories are the
 models for the Hodge theory [18, 30-32]. \\

 \noindent
{\bf Acknowledgments:}\\

\noindent
The present investigation has been carried under the BHU-fellowship received by S. Kumar and
DST-INSPIRE fellowship (Govt. of India) awarded to B. Chauhan. Both these authors express their gratefulness to
the above funding agencies for the financial supports.\\

\noindent
{\bf Declaration:}\\

\noindent
The authors declare that there is no conflict of interests of any kind.\\

\begin{center}
{\bf Appendix A: On the Derivation of (Anti-)BRST Symmetries}\\
\end{center}

\noindent
Here we derive the (anti-)BRST symmetries (cf. Sec. 2) by exploiting the simple 
(but fruitful) augmented version of (anti-)chiral superfield approach (ACSA) to
 BRST formalism [30-32] where the (anti-)BRST invariant restrictions
play very crucial roles. In this context, first of all, we generalize the basic 2D fields (e.g. $A_\mu$, $C$, $\bar C$) 
onto $(2, 1)$-dimensional anti-chiral super-submanifold (of the general (2, 2)-dimensional supermanifold) as:
\[A_\mu(x)\to B_\mu (x,\bar\theta) = A_\mu (x) + \bar \theta \, R_\mu (x),\] 
\[C(x)\to F(x,\bar\theta) = C(x) + i\, \bar\theta \,B{_1}(x),\] 
\[\bar C(x) \to \bar F(x,\bar\theta) = \bar C(x) + i\, \bar \theta \, B{_2}(x), \eqno (A.1)\]
which are nothing but the limiting cases of the  super expansions
 in Eq. (10) (that are on the general (2, 2)-dimensional supermanifold). 
It is worthwhile to mention here that the
Nakanishi-Lautrup auxiliary field $B(x)$ has {\it no} anti-chiral expansion (i.e. $B(x)\rightarrow \tilde B(x,\bar\theta)
 = B(x))$ because we note that $s_b B(x) = 0$. We further point out that $s_b (\bar C\times B) = 0$.
 This observation can be generalized onto the anti-chiral $(2, 1)$-dimensional super-submanifold with
the following restriction on the superfields due to the ACSA to BRST formalism:
\[\bar F(x,  \bar\theta) \times\tilde B(x,\bar\theta) = \bar C (x)\times B(x),\qquad\qquad \tilde B(x,\bar\theta) = B(x),\eqno (A.2)\] 
which leads to $B_2 \times B = 0$. One of the non-trivial solutions is $B_2$ is proportional to $B $.
For the sake of brevity, however, we choose $B_2 = B$.
The above restriction (A.2) is consistent with the basic tenets of AVSA/ACSA to BRST 
formalism where we demand that the BRST invariant quantity  should be independent of 
$\bar\theta$ variable.
Thus, we have 
\[\bar F^{(b)}(x,\bar\theta) = \bar C(x) + i\, \bar \theta \, B(x)\equiv \bar C(x) + \bar\theta \,  (s_b \bar C), \eqno (A.3)\]
where the superscript $(b)$ denotes that the superfield $\bar F^{(b)} (x, \bar\theta)$ has been obtained after the application of  BRST
invariant restriction (A.2). It goes without saying that (in the 
above process) we have derived
the BRST transformation for $\bar C $ as: $s_b \bar C = iB$ (cf. Sec. 2).

We carry out the above kinds of exercises to obtain the  {\it other} BRST symmetry transformations
associated with the other fields of the theory. In this context, first of all, we observe 
that the following are the useful BRST-invariant quantities (in
addition to the earlier BRST-invariant quantities: $s_b B = 0,\;s_b(B\times\bar C) = 0$), namely;
\[s_b (D_{\mu} C) = 0,\qquad s_b (C\times C) = 0,\]
\[s_b (A^{\mu}\cdot\partial_{\mu} B + i\; \partial_{\mu}\bar C\cdot D^{\mu}C) = 0.\eqno (A.4)\]
According to the  AVSA/ACSA to BRST formalism, the above quantities 
can be generalized onto the (2, 1)-dimensional anti-chiral super-submanifold and the corresponding superfields can be restricted 
to obey the following conditions:
\[\partial_{\mu} F (x,\bar\theta) + i\; \big (B_{\mu}(x,\bar\theta)\times F(x,\bar\theta)\big) =
\partial_{\mu} C(x) + i\; (A_{\mu}(x)\times C(x)),\]
\[F(x,\bar\theta)\times F(x,\bar\theta) = C(x)\times C(x),\]
\[B^{\mu}(x,\bar\theta)\cdot \partial_{\mu} B(x) + i\; \partial_{\mu} \bar F^{(b)}(x,\bar\theta)\cdot
\partial^{\mu} F(x,\bar\theta) - \partial_{\mu}\bar F ^{(b)}(x,\bar\theta)\cdot \big(B^{\mu}(x,\bar\theta)\times F(x,\bar\theta)\big)\]
\[~~~~~~~~~~~~~~~~~~~~~ = A^{\mu}(x)\cdot\partial_{\mu} B(x) + i\;\partial _{\mu}\bar C(x)\cdot \partial^{\mu} C(x) -
\partial_{\mu}\bar C(x)\cdot (A^{\mu}(x)\times C(x)).\eqno (A.5)\]
In other words, we demand that the l.h.s. of the above equality should remain 
independent  of {\it ``soul''} coordinate $\bar\theta$.
The above requirements lead to the following:
\[D_{\mu} B_1 (x) + R_{\mu}(x)\times C(x) = 0,\;~~~~~~~~ B_1(x)\times C(x) = 0\]
\[R^{\mu}(x)\cdot\partial_{\mu} B(x) + \partial_{\mu}\bar C(x)\cdot(R^{\mu}(x)\times C(x)) + \partial_{\mu}\bar C(x)
\cdot D^{\mu} B_1 (x) - \partial _{\mu} B(x)\cdot D^{\mu} C(x) = 0.\eqno (A.6)\]
We discuss here the solutions of the above conditions. It is clear that $ B_1 (x )$ is proportional
to $(C(x)\times C(x))$ because we have obtained the condition $B_1(x)\times C(x) = 0$ in (A.6).
Thus, the non-trivial expression for  $B_1(x) = \kappa \; (C(x)\times C(x))$ where $\kappa $ is
a numerical constant. From the relation $D_{\mu} B_1(x) + R_{\mu}(x)\times C = 0$,
it is clear that the following choices
\[B_1(x) = - \frac {1}{2} (C(x)\times C(x)),\qquad\qquad R_{\mu}(x) = D_{\mu}C(x),\eqno (A.7)\]
satisfy the relation $B_1(x)\times C(x) = 0$ and $D_{\mu}B_1 + (R_{\mu}\times C) = 0 $ {\it together}.
It is gratifying to note that these conditions also satisfy the last relationship that has been
quoted in (A.6). Thus, ultimately, we have obtained the following expansions (with superscript $(b)$):
\[B_{\mu}^{(b)}(x,\bar\theta) = A_{\mu}(x) + \bar\theta\; (D_{\mu} C)\equiv A_{\mu}(x) +\bar\theta\;(s_b A_{\mu}(x)),\]
\[F^{(b)}(x,\bar\theta) = C(x) +\bar\theta\; [- \frac {i}{2}\; (C\times C)]\equiv C(x) + \bar\theta\; (s_b C(x)),\]
\[\bar F^{(b)}(x,\bar\theta) = \bar C(x) + \bar\theta\; (i\; B(x))\equiv \bar C(x) + \bar\theta\; (s_b\bar C(x)).\eqno (A.8)\]
In other words, we have already derived the BRST symmetry transformations $s_b$ for the basic fields $(A_{\mu}, C, \bar C)$
which are nothing but the coefficients of $\bar\theta$ in the superfields expansions (A.8). The sanctity of this 
statement can be checked from Eq. (2).

Finally, we comment on the derivation of the BRST symmetry transformations:
$s_b {\cal B} = i\;({\cal B}\times C),~~~ s_b E = i\;(E\times C)$ and $s_b \bar B = i\,(\bar B\times C).$
In this context, we have the following generalizations on the (2, 1)-dimensional anti-chiral super-submanifold:
\[{\cal B}(x)\longrightarrow \tilde {\cal B}(x,\bar\theta) = {\cal B}(x) + \bar\theta\; P(x),\]
\[E(x)\longrightarrow \tilde E (x,\bar\theta) = E(x) + \bar\theta\; Q(x),\]
\[\bar B(x)\longrightarrow \tilde{\bar B}(x,\bar\theta) = \bar B(x) + \bar\theta\; S(x),\eqno (A.9)\]
where $(P(x), Q(x), S(x))$ are the fermionic
secondary fields which have to be determined in terms of the basic and 
auxiliary fields of the theory. We note the following:
\[s_b ({\cal B}\times C) = 0\qquad s_b (E\times C) = 0,\qquad s_b(\bar B\times C) = 0.\eqno (A.10)\]
According to the basic tenets of AVSA/ACSA, we have 
the following equalities 
\[\tilde {\cal B}(x,\bar\theta)\times F^{(b)}(x,\bar\theta) = {\cal B}(x)\times C(x),\]
\[\tilde E(x,\bar\theta)\times F^{(b)}(x,\bar\theta)= E(x)\times C(x),\]
\[\tilde{\bar B}(x,\bar\theta)\times F^{(b)}(x,\bar\theta) = \bar B(x)\times C(x),\eqno (A.11)\]
which show that the BRST invariant quantities of (A.10) should remain independent
of the ``soul'' coordinate $\bar\theta$. This restriction yields the following:
 \[P(x) = i\; ({\cal B}\times C),\qquad Q(x) = i\; (E\times C),\qquad S(x) = i\;(\bar B\times C).\eqno (A.12)\]
 Thus, ultimately, we have derived the following:
 \[\tilde{\cal B}^{(b)}(x,\bar\theta) = {\cal B}(x) + \bar\theta \;[ i\;({\cal B}\times C)]\equiv {\cal B}(x) +\bar\theta\; (s_b {\cal B}(x)),\]
 \[\tilde E^{(b)}(x,\bar\theta) = E(x) + \bar\theta\; [ i \;(E\times C)]\equiv E(x) +\bar\theta\; (s_b E(x)),\]
 \[ \tilde{\bar B}^{(b)}(x,\bar\theta) = \bar B(x) + \bar\theta\; [ i\; (\bar B\times C)]\equiv \bar B(x) +\bar\theta \;(s_b \bar B(x)),\eqno (A.13)\]
where superscript $(b)$ denotes the fact that the above super expansions have been derived 
after the application of the BRST invariant restrictions (A.10) and (A.11).
From Eq. (A.13), we note that coefficients of $\bar\theta$ are nothing but
the BRST symmetry transformations for $ {\cal B},\; E\;$ and $ \bar B $ fields as 
given in Eq. (2) (cf. Sec. 2 for details).

We now focus on the derivation of the anti-BRST symmetry by chiral superfield approach to BRST 
formalism where we have the following  generalizations:
\[A_\mu(x)\to B_\mu (x,\theta) = A_\mu (x) +  \theta \,\bar R_\mu (x),\] 
\[C(x)\to F(x,\theta) = C(x) + i\, \theta \,\bar B{_1}(x),\] 
\[\bar C(x) \to \bar F(x,\theta) = \bar C(x) + i\,  \theta \,\bar B{_2}(x), \eqno (A.14)\]
 where $(\bar R_{\mu},\; \bar B_1,\;\bar B_2)$ are the secondary fields that
have to be determined in terms of the basic and auxiliary fields of the theory by invoking the 
anti-BRST invariant restrictions. It goes without saying that the above expansions are the limiting cases 
of the super expansions in Eq. (10) when $\bar\theta = 0$.
It can be checked that we have the following useful and interesting  
anti-BRST invariant quantities (cf. Sec. 2):
\[s_{ab}\bar B  = 0,\qquad s_{ab} (\bar B\times C) = 0,\qquad s_{ab}(\bar C\times \bar C) = 0,\]
\[s_{ab}[A^{\mu}\cdot\partial_{\mu}\bar B - i D_{\mu}\bar C\cdot \partial^{\mu} C ] = 0,\qquad
s_{ab} (D_{\mu}\bar C) = 0.\eqno (A.15)\]
According to the basic  tenets of AVSA/ACSA, we have to demand that the above 
quantities (when generalized onto (2, 1)-dimensional chiral supermanifold)
should be independent of the Grassmannian variable $\theta$.
In other words, we have the following equalities:
\[\bar B (x)\times F(x,\theta) = \bar B (x)\times C(x),\qquad \bar F(x,\theta)\times\bar F(x,\theta) =
\bar C(x)\times \bar C(x),\]
\[\partial_{\mu}\bar F(x,\theta) + i\; B_{\mu}(x,\theta)\times\bar F(x,\theta) = \partial_{\mu}\bar C(x)
+ i\, (A_{\mu}(x)\times \bar C(x)),\]
\[B^{\mu}(x,\theta)\cdot\partial_{\mu}\bar B(x) - i\; \partial _{\mu}\bar F(x,\theta)\cdot\partial ^{\mu} F(x,\theta)
+ (B_{\mu}(x,\theta)\times \bar F(x,\theta))\cdot\partial^{\mu}F(x,\theta)\]
\[~~~~~~~~~~~~~~~~~~~~~~~~~~~~~~~~~~~~~= A^{\mu}(x).\partial_{\mu}\bar B(x) - 
i\;D_{\mu}\bar C(x)\cdot \partial^{\mu} C(x).\eqno(A.16)\]
We note here that, because of $s_{ab}\bar B = 0$, we have {\it no} chiral super expansion of $\bar B(x)$ 
(i.e. $\bar B(x) \longrightarrow \tilde{\bar B}(x,\theta) =\bar B(x))$.
The above equalities lead to the following expressions for the secondary 
fields $(\bar R_{\mu},\bar B_1,\bar B_2)$ in terms of basic and auxiliary fields of our theory:
\[ \bar R_{\mu} = D_{\mu}\bar C,\qquad \bar B_1 =\bar B,\qquad \bar B_2 = -\frac{1}{2}(\bar C\times\bar C).\eqno (A.17)\]
 Thus, we have obtained the following chiral super expansions
 \[B_{\mu}^{(ab)}(x,\theta) = A_{\mu}(x) + \theta\; (D_{\mu}\bar C)\equiv A_{\mu}(x) + \theta\; (s_{ab} A_{\mu}),\]
 \[F^{(ab)}(x,\theta) = C(x) + \theta \;(i\;\bar B)\equiv C(x) + \theta\; (s_{ab} C),\]
 \[\bar F^{(ab)}(x,\theta) = \bar C(x) + \theta\; [ -\;\frac{i}{2}\;(\bar C\times \bar C)]\equiv 
 \bar C(x) +\theta\; (s_{ab}\bar C),\eqno (A.18)\]
 where the  superscript $(ab)$ denotes the super expansions of the chiral superfields after the application
 of the anti-BRST invariant restrictions [cf. (A.15), (A.16)]. A close look at (A.18)
 demonstrates that we have already obtained the anti-BRST symmetry transformations (cf. Sec. 2)
for the basic fields $A_{\mu}(x), C(x)$ and $\bar C(x)$ of our theory.

 Now we dwell a bit on the derivation of
the anti-BRST symmetry transformations:
 $s_{ab} B = i\; (B\times\bar C)$, $s_{ab} E = i\; (E\times\bar C)= 0$
 and $s_{ab} {\cal B} = i\; ({\cal B}\times\bar C)$.
 In this connection, we note that the following are the useful anti-BRST invariant 
 quantities for our further discussion:
 \[s_{ab}(B\times\bar C) = 0,\qquad s_{ab}(E\times\bar C) = 0,\qquad s_{ab}({\cal B}\times \bar C) = 0.\eqno (A.19)\]
 According to the basic principles  of AVSA/ACSA, the above quantities should be independent of 
 the Grassmannian variable $\theta$ when they are generalized onto the (2, 1)-dimensional chiral super-submanifold.
 In other words, we have the following equalities 
 \[\tilde {\cal B}(x,\theta)\times\bar F^{(ab)}(x,\theta) = B(x)\times\bar C(x),\]
 \[\tilde E(x,\theta)\times\bar F^{(ab)}(x,\theta) = E(x)\times\bar C(x),\]
 \[\tilde{\cal B}(x,\theta)\times\bar F^{(ab)}(x,\theta) \equiv {\cal B}(x)\times\bar C(x),\eqno (A.20)\]
 where the expansion for the chiral superfield $\bar F^{(ab)}(x,\theta)$ has been given in (A.18) and the chiral
 super expansions of the other superfields are as follows:
 \[B(x)\longrightarrow \tilde B (x,\theta) = B(x) + \theta\;\bar P(x),\]
 \[E(x)\longrightarrow \tilde E (x,\theta) = E(x) + \theta\;\bar Q(x),\] 
 \[{\cal B}(x)\longrightarrow \tilde{\cal B} (x,\theta) = {\cal B}(x) + \theta\;\bar S(x). \eqno (A.21)\]
 Hence, the fields $(\bar P(x), \bar Q(x), \bar S(x))$ are the fermionic secondary fields that are to be determined in terms of the basic and auxiliary 
 fields of our 2D non-Abelian theory from the anti-BRST invariant restrictions (cf. (A.19), (A.20)). Explicit substitution 
 of expansions from (A.18) and (A.21) lead to the following very useful and interesting  relationships:
\[\bar P(x) = i\;(B(x)\times\bar C(x)), \qquad \bar Q(x) = i\;(E(x)\times\bar C(x)),\qquad \bar S(x) = i\;({\cal B}(x)\times\bar C(x)). \eqno (A.22)\]
These relationships prove the fermionic nature of the secondary fields $(\bar P(x), \bar Q(x), \bar S(x))$
which is also evident from (A.21) due to the fermionic $(\theta^2 = 0)$ nature of $\theta$.
 Thus, we have the following super expansions for the superfields in (A.21), namely;
 \[ \tilde{B}^{(ab)}(x,\bar\theta) = B(x) + \theta\; [ i\; (B\times\bar C)]\,\equiv \, B(x) + \theta \;(s_{ab} B(x)),\]
 \[\tilde E^{(ab)}(x,\bar\theta) = E(x) + \theta\; [ i \;(E\times\bar C)]\,\equiv \,E(x) + \theta\; (s_{ab} E(x)),\]
 \[\tilde{\cal B}^{(ab)}(x,\bar\theta) = {\cal B}(x) + \theta \;[ i\;({\cal B}\times\bar C)]\,\equiv\, {\cal B}(x) + \theta\; (s_{ab} {\cal B}(x)), \eqno (A.23)\]
 where the superscript $(ab)$ denotes the super expansions of the superfields after the application of the anti-BRST invariant restrictions (A.20).
 The coefficients of $\theta$ in (A.23) are nothing but the anti-BRST symmetry transformations for the
 fields $B(x), E(x)$ and ${\cal B}(x)$. Thus, we have derived {\it all} the (anti-)BRST symmetry transformations 
of our non-Abelian theory by applying the (anti-)chiral superfield approach to BRST formalism.

\begin{center}
{\bf Appendix B: On the Derivation of (Anti-)co-BRST Symmetries}\\
\end{center}
We derive here the nilpotent and absolutely anticommuting (anti-)co-BRST symmetry transformations by exploiting 
the virtues of the (anti-)co-BRST invariant restrictions within the framework of the (anti-)chiral 
superfield approach to BRST formalism. In this context, first of all, we take the anti-chiral super expansions 
(A.1) as well as (A.9) and focus on the following
very useful and interesting co-BRST invariant quantities:
\[s_d \bar C = 0, \quad s_d (\partial_{\mu}A^{\mu}) = 0, \quad s_d {\cal B} = 0, \quad s_d (D_{\mu}\partial^{\mu}\bar C) = 0,
\quad s_d B = 0, \quad s_d\bar B = 0,\]
\[ s_d (C\times {\cal B}) = 0,\qquad s_d[\varepsilon ^{\mu\nu}A_{\nu} \cdot \partial_{\mu}{\cal B} - i\;\partial_{\mu}\bar C \cdot \partial^{\mu} C] = 0. \eqno (B.1)\]
It is crystal clear that the co-BRST invariant quantities, 
when generalized onto the $(2, 1)$-dimensional anti-chiral 
super-submanifold
(of the general $(2, 2)$-dimensional supermanifold) should be
 independent of the Grassmannian coordinate $\bar\theta$.
Against this backdrop, it is very edident, (from (A.1) and (A.9)), 
that the following are true:
\[\bar C(x)\longrightarrow \bar F^{(d)}(x, \bar\theta) = \bar C(x) 
+\bar\theta\;(0)\;\Longrightarrow\;  B_2 = 0,\qquad s_d\bar C = 0,\]
\[~~{\cal B}(x)\longrightarrow \tilde {\cal B}^{(d)}(x, \bar\theta)
 = {\cal B}(x)+\bar\theta\;(0)\;\Longrightarrow\;  P(x) = 0,\qquad s_d {\cal B} = 0,\]
\[~~~~~~~~~~~~~~\bar B(x)\longrightarrow \bar B^{(d)}(x, \bar\theta) = \bar B(x)
+\bar\theta\;(0)\;\Longrightarrow\;  S(x) = 0,\qquad s_d \bar B(x) = 0, \eqno (B.2)\]
 where the superscript $(d)$ on the superfields denotes that 
the above superfields have been derived 
after the application of co-BRST invariant restrictions (B.1) 
which demonstrate that the 
co-BRST invariant quantities should be independent of the soul 
coordinate $\bar\theta$
(due to the basic tenets of augmented version of (anti-)chiral superfield 
approach to BRST formalism). Further,
the other co-BRST invariant quantities in (B.1) imply
\[\partial_{\mu}\partial^{\mu}\bar C(x) 
+ i B_{\mu}(x,\bar\theta)\cdot\partial^{\mu}\bar C(x) =
 D_{\mu}\partial^{\mu}\bar C(x)\Longrightarrow 
R_{\mu}\times\partial^{\mu}\bar C = 0,\]
\[\partial_{\mu} B^{\mu}(x,\bar\theta) 
= \partial_{\mu}A^{\mu}(x)\Longrightarrow \partial_{\mu}R^{\mu} = 0.\eqno(B.3)\]
It is evident that the non-trivial co-BRST 
symmetry transformations are $s_d C = -i {\cal B}$
 and 
$s_d A_{\mu} = -\varepsilon_{\mu\nu}\partial^{\nu}\bar C$. 
 These can be derived from the co-BRST invariant
restrictions:
\[\tilde{\cal B}^{(d)}(x,\bar\theta)\times F(x,\bar\theta) = {\cal B}(x)\times C(x),\]  
\[\varepsilon^{\mu\nu} B_{\nu}(x,\bar\theta)\cdot\partial_{\mu} {\cal B}^{(d)}(x,\bar\theta)
- i\;\partial_{\mu}\bar F^{(d)}(x,\bar\theta)\cdot\partial^{\mu}F^{(d)}(x,\bar\theta)\]
\[= \varepsilon^{\mu\nu} A_{\nu}(x)\cdot\partial_{\mu}{\cal B}(x) -
 i\;\partial_{\mu} \bar C(x)\cdot\partial^{\mu}C(x).\eqno (B.4)\]
 The substitution of the expansion of $F(x,\bar\theta)$ 
from (A.1) into the top relationship, in the above,
 leads to the condition $ B_1\times{\cal B} = 0$. 
One of the non-trivial solution is $ B_1 = - {\cal B} $
 so that we obtain the following useful expansion:
 \[F^{(d)}(x,\bar\theta) = C(x) + \bar\theta \;(-i\;{\cal B})
\equiv C(x) + \bar\theta\;(s_d C). \eqno(B.5)\]
Finally, when we substitute the expansions for $B_{\mu}(x,\bar\theta)$
 from (A.1),  ${\cal B}^{(d)}(x,\bar\theta),
 \bar F^{(d)}(x,\bar\theta)$ from (B.2) and $F^{(d)}(x,\bar\theta)$
 from (B.5), we obtain:
 $R_{\mu}(x) = -\varepsilon_{\mu\nu}\partial^{\nu}\bar C$ 
which {\it also} satisfies both the
additional conditions in  (B.3) and leads to: 
\[B_{\mu}^{(d)}(x,\bar\theta) = A_{\mu}(x) + \bar\theta\;
(-\varepsilon_{\mu\nu}\partial^{\nu}\bar C)\equiv 
A_{\mu}(x) +\bar\theta \;(s_d A_{\mu}).\eqno (B.6)\]
The super expansions in (B.2), (B.5) and (B.6) demonstrate
 that we have derived:
$s_d \bar B = s_d {\cal B} = s_d \bar C = 0,~~ s_d C = - 
i {\cal B},~~ s_d A_{\mu} = -\varepsilon_{\mu\nu}\partial^{\nu}\bar C$.
We mention, in passing, that $s_d B = 0$ implies that  we have {\it no}
anti-chiral expansion for $B(x)$ as it is a co-BRST invariant quantity.

To derive the anti-co-BRST symmetry transformations,
 we invoke the chiral expansions for the superfields as
given in (A.14) and (A.20). In this context,
 first of all, we look for the useful anti-co-BRST
invariant quantities and generalize them onto 
(2, 1)-dimensional {\it chiral} super submanifold (of the 
general (2, 2)-dimensional supermanifold on which 
our present theory is generalized). 
After this, we demand that such invariant 
quantities should be independent of 
the ``soul'' coordinate $\theta$. In this context, we note the following:
\[ s_{ad} C = 0,\qquad s_{ad}(\partial_{\mu}A^{\mu}) = 
0,\qquad s_{ad} (D_{\mu}\partial^{\mu} C) = 0,
\qquad s_{ad}\bar B = 0,\]
\[s_{ad}(\bar C\times {\cal B}) = 0,\qquad
 s_{ad} B = 0,\qquad  s_{ad} {\cal B} = 0,\]
\[s_{ad}[\varepsilon^{\mu\nu}A_{\nu}\cdot\partial_{\mu}{\cal B}-
 i\;\partial_{\mu}\bar C\cdot\partial^{\mu} C] = 0.\eqno (B.7)\] 
 The {\it trivial} chiral expansions are: 
$ B(x)\longrightarrow \tilde B^{(ad)}(x,\theta) = B(x),~~~
 \bar B(x)\longrightarrow \tilde{\bar B}^{(ad)}(x,\theta) =\bar B(x),
 {\cal B}(x)\longrightarrow \tilde{\cal B}^{(ad)}(x,\theta) =
 {\cal B}(x),~~~ C(x)\longrightarrow F^{(ad)}(x,\theta) = C(x)$ 
which imply $s_{ad} C = s_{ad} B = s_{ad}\bar B = s_{ad}{\cal B} = 0$.
The non-trivial conditions are:
\[\partial_{\mu}B^{\mu}(x,\theta) =\partial_{\mu}A^{\mu}(x)
\Longrightarrow \partial_{\mu}\bar R^{\mu} = 0,\]
\[\bar F(x,\theta)\times {\cal B}^{(ad)}(x,\theta) = 
\bar C(x)\times{\cal B}(x)\Longrightarrow (\bar B_2\times {\cal B}) = 0,\eqno (B.8)\]
which imply that if we choose $\bar B_2 = {\cal B}$, 
the condition $\bar B_2\times {\cal B} = 0$ is satisfied 
and it leads to 
\[\bar F^{(ad)}(x,\theta) = \bar C(x) + 
\theta (i\,{\cal B})\equiv \bar C(x) + \theta (s_{ad}\bar C),\eqno (B.9)\]
where the superscript $(ad)$ denotes that 
the above superfield has been obtained after the application of (B.7).
Thus, we observe that we have already derived the non-trivial anti-co-BRST symmetry transformation:
$s_{ad}\bar C = i {\cal B}$. We now focus 
on the latter conditions
\[\partial_{\mu}\partial^{\mu} F^{(ad)}(x,\theta) +
 i\; B_{\mu}(x,\theta)\times\partial^{\mu} F^{(ad)}(x,\theta)=
\partial_{\mu}\partial^{\mu}C(x) + i\; A_{\mu}(x)\times \partial^{\mu} C(x),\]
\[\varepsilon^{\mu\nu}B_{\nu}(x,\theta)\cdot\partial_{\mu}{\cal B}(x)-
 i \;\partial_{\mu}\bar F^{(ad)}(x,\theta)\cdot
\partial^{\mu}F^{(ad)}(x,\theta)\] 
\[= \varepsilon^{\mu\nu}A_{\nu}(x)\cdot\partial_{\mu}{\cal B}(x) - i\; \partial_{\mu}\bar C(x)
\cdot\partial^{\mu} C(x),\eqno (B.10)\]
where we have to use $F^{(ad)}(x,\theta) = C(x)$ and (B.9) to obtain the following 
conditions:
\[\bar R_{\mu}\times\partial^{\mu} C = 0,\qquad \bar R_{\mu} +
 \varepsilon_{\mu\nu}\partial^{\nu} C =0\Longrightarrow \bar R_{\mu}
 = -\varepsilon_{\mu\nu}\partial^{\nu} C.\eqno (B.11)\]
 Thus, ultimately, we obtain the chiral expansion:
 \[B_{\mu}^{(ad)}(x,\theta) = A_{\mu}(x) + \theta \;(-\varepsilon_{\mu\nu}\partial^{\nu}C)\equiv 
 A_{\mu}(x) +\theta\; (s_{ad} A_{\mu}),\eqno (B.12)\]
 where the supersript $(ad)$ denotes that the above
 superfield has been obtained after the application of (B.7).
 It is evident, by now, that we have obtained {\it all}
 the anti-co-BRST symmetry transformations 
 $s_{ad}$ (cf. Sec. 2) of our theory by exploiting the
 symmetry invariant  restrictions
 on the {\it chiral} superfields. We point out that the choice
 $R_{\mu}  = - \varepsilon_{\mu\nu}\partial^{\nu}C$
 satisfies both the  additional conditions
 $\partial_{\mu}\bar R^{\mu} = 0$ and $\bar R_{\mu}\times \partial^{\mu} C = 0$
 which are present in (B.9) and (B.11). We comment that
 we have chosen $B_1 = - {\cal B}$ and $\bar B_2 = +{\cal B}$
 (which imply $s_d C  = - i {\cal B}$ and $s_{ad}\bar C = i {\cal B}$)
 because these choices satisfy the absolute 
anticommutativity property ($s_d s_{ad}+ s_{ad} s_d = 0$) of 
the (anti-)co-BRST symmetry transformations.

 \begin{center}
{\bf Appendix C: On the Symmetry Invariance in the Theory}\\
\end{center}
We have concentrated on the (anti-)BRST as well as (anti-)co-BRST invariance(s) of our present
2D non-Abelian theory within the framework of AVSA to 
BRST formalism. In this Appendix, we capture the (anti-)BRST and (anti-)co-BRST
invariance of the Lagrangian densities (1) and (39)
within the framework of AVSA to BRST formalism 
(which are explicitly quoted in Eqs. (3), (7) and (40)). 
Towards this goal in mind, first of all, we generalize 
the Lagrangian densities (1) onto (2, 2)-dimensional
 supermanifold  as follows,

\[{\cal L}_B \longrightarrow \tilde {\cal L}_B = {\cal B}^{(g)}(x,\theta,\bar\theta)\cdot \tilde E^{(h)}(x,\theta,\bar\theta)
- \frac {1}{2}\; {\cal B}^{(g)}(x,\theta,\bar\theta)\cdot  {\cal B}^{(g)}(x,\theta,\bar\theta) 
 + B(x)\cdot\partial_{\mu}B^{\mu(h)}(x,\theta,\bar\theta)\]
\[+ \frac {1}{2}\; (B(x)\cdot B(x) 
+ \tilde {\bar B}^{(g)}(x,\theta,\bar\theta)\cdot\tilde {\bar B}^{(g)}(x,\theta,\bar\theta))
- i\; \partial_{\mu}\bar F^{(h)}(x, \theta, \bar\theta)\cdot D^{\mu} F^{(h)}(x,\theta,\bar\theta),\]
 \[{\cal L}_{\bar B} \longrightarrow \tilde {\cal L}_{\bar B} 
= {\cal B}^{(g)}(x,\theta,\bar\theta)\cdot \tilde E^{(h)}(x,\theta,\bar\theta)
- \frac{1}{2}\; {\cal B}^{(g)}(x,\theta,\bar\theta)\cdot  {\cal B}^{(g)}(x,\theta,\bar\theta) 
 - \bar B(x)\cdot\partial_{\mu}B^{\mu(h)}(x,\theta,\bar\theta)\]
\[+ \frac {1}{2}\; (B^{(g)}(x,\theta,\bar\theta)\cdot B^{(g)}(x,\theta,\bar\theta)) + \frac {1}{2}\; \bar B(x)\cdot\bar B(x)
- i\; D_{\mu}\bar F^{(h)}(x,\theta,\bar\theta)\cdot\partial^{\mu} F^{(h)}(x,\theta,\bar\theta),\eqno (C.1) \]
where the superfields with superscript $(h)$ are the ones that have been derived in the main body of the 
text. It is to be noted that we have defined the covariant derivatives as:
$D_{\mu} F^{(h)}(x,\theta,\bar\theta) =
 \partial_{\mu}F^{(h)}(x,\theta,\bar\theta) 
+ i\; (B_{\mu}^{(h)}(x,\theta,\bar\theta)\times F^{(h)}
(x,\theta,\bar\theta))$ and $D_{\mu}\bar F^{(h)}(x,\theta,\bar\theta)
 = \partial_{\mu}\bar F^{(h)}(x,\theta,\bar\theta) 
+ i\; (B_{\mu}^{(h)}(x,\theta,\bar\theta)\times \bar F^{(h)}
(x,\theta,\bar\theta))$. The superfields with superscript $(g)$ denote the {\it ones} that have been obtained after GIR. We elaborate 
here a few of them. For instance, let us focus on the explicit expression of ${\cal B}^{(g)}(x, \theta, \bar\theta)$.
In this context, we note that:
\[s_b(E\cdot {\cal B}) = 0,~~~~~~s_{ab}(E\cdot {\cal B}) = 0.\eqno (C.2)\]
At this stage, we exploit the basic tenets of AVSA to  BRST formalism which state that any arbitrary 
(anti-)BRST invariant quantity must remain independent of the ``soul'' coordinates $(\theta, \bar\theta)$ when it is 
generalized onto an appropriately chosen supermanifold on which our basic gauge theory is generalized.
Thus, we have the following equality: 
\[\tilde E^{(h)}(x, \theta, \bar\theta)\cdot {\cal B} (x, \theta, \bar\theta) = E(x)\cdot{\cal B}(x).\eqno (C.3)\]
In the above, the full expansions for $E^{(h)}(x, \theta, \bar\theta)$ and ${\cal B}(x, \theta, \bar\theta)$
are
\[E^{(h)}(x, \theta, \bar\theta) = E(x) + \theta\; (i\; E\times\bar C) + \bar\theta\; ( i\; E\times C)
+ \theta\; \bar\theta\; [ - E\times B - (E\times C)\times\bar C],\]
\[{\cal B}(x, \theta, \bar\theta) = {\cal B}(x) + \theta \;\bar S(x) + \bar\theta\; S(x) + i\;\theta\;\bar\theta\; P(x),\eqno (C.4)\]
where $E^{(h)}(x, \theta, \bar\theta)$ has been derived from Eq. (18) and the general super expansion for the 
superfield ${\cal B}(x, \theta, \bar\theta)$ has been quoted in (C.4) where the secondary fields
$(S(x), \bar S(x))$ are fermionic and $P(x)$ is bosonic in nature. The substitution of (C.4) into
(C.3) produces the following expressions for the secondary fields in terms of the basic and auxiliary fields:
\[ S(x) = i\; ({\cal B}\times C),~~~~~~\bar S(x) = i\; ({\cal B}\times\bar C),\]
\[P(x) = i\; \Big [\big({\cal B}\times B\big) + \big ({\cal B}\times C\big)\times\bar C\Big].\eqno (C.5)\]
Thus, we have the final expansion for the superfield ${\cal B}^{(g)}(x, \theta, \bar\theta)$
as:
\[{\cal B}^{(g)}(x, \theta, \bar\theta) = {\cal B}(x) + \theta\,(i\; {\cal B}\times\bar C)
+\bar\theta \,(i\; {\cal B}\times C) + \theta\;\bar\theta\; \Big [-\;{\cal B}\times B -\;({\cal B}\times C)\times\bar C\Big],\]
\[\equiv  {\cal B}(x) + \theta\,(s_{ab}\; {\cal B}) + \bar \theta\,(s_{b}\; {\cal B}) + \theta\;\bar\theta \,(s_b \; s_{ab}\; {\cal B}).\eqno (C.6)\]
In other words, we have derived the (anti-)BRST symmetry transformations for the auxiliary field ${\cal B}(x)$ and, in the process, we have obtained 
the explicit form of ${\cal B}^{(g)} (x, \theta, \bar\theta)$ which has been used in the explicit 
expression for the super Lagrangian densities (C.1).
We discuss here about the derivations of $\tilde{\bar B}^{(g)}(x,\theta,\bar\theta)$
and $\tilde B^{(g)}(x,\theta,\bar\theta)$ that are present in the expressions for 
the super Lagrangian densities $\tilde{\cal L}_B$ and $\tilde {\cal L}_{\bar B}$
(cf. (C. 1)). Using the (anti-)BRST symmetry transformations from Eq. (2), we note that 
the following
\[ s_b (E\cdot \bar B) = 0,\qquad s_{ab} (E\cdot B) = 0,\qquad
s_{ab}\bar B = 0,\qquad s_b B = 0,\eqno (C.7)\]
are the BRST and anti-BRST invariant quantities. According to the basic tenets of AVSA
to BRST formalism, the BRST invariance of $B$ (i.e. $s_b B = 0$) and anti-BRST invariance of 
$\bar B$ (i.e. $s_{ab}\bar B = 0$) imply that the following general super expansions
\[B(x)\quad\longrightarrow\quad\tilde B(x,\theta,\bar\theta) = B(x) + \theta\;\bar M(x)
 +\bar\theta\;M(x) + i\;\theta\;\bar\theta\; N(x),\]
\[\bar B(x)\quad\longrightarrow\quad \tilde {\bar B}(x,\theta,\bar\theta) = \bar B(x) 
+ \theta\;\bar L(x) + \bar\theta \, L(x) + i \theta\;\bar\theta\; K(x),\eqno (C.8)\] 
would remain independent of $\bar\theta$ and $\theta$, 
respectively, in
view of the mapping $s_b\longleftrightarrow \partial_{\bar\theta}$ 
and $s_{ab}\longleftrightarrow \partial_{\theta}$.
Thus, the reduced form of the superfields in (C.8) are:
\[\tilde B^{(r)}(x,\theta,\bar\theta) = B(x) + \theta\;\bar M(x), \qquad
\tilde{\bar B}^{(r)}(x,\theta,\bar\theta) = \bar B(x) + \bar\theta\;\bar L(x).\eqno (C.9)\] 
 In the above expansions (C.8) and (C.9), the secondary fields
 $(M(x), \bar M(x), L(x),\bar L(x))$ are fermionic and $(N(x), K(x))$ 
are bosonic in nature due to the 
fermionic nature (i.e. $\theta^2 = \bar\theta^2 = 0, 
\theta\;\bar\theta + \bar\theta\;\theta = 0$)
of the Grassmannian variables $(\theta,\bar\theta)$ and bosonic 
nature of the superfields $\tilde B(x,\theta,\bar\theta)$
and   $\tilde {\bar B}(x,\theta,\bar\theta)$. 
The superscript $(r)$ on the superfields in (C.9) corresponds
to the reduced form of the general super expansion in (C.8) 
when $\bar\theta = 0$ and $\theta = 0$, respectively. Basically, these reduced forms become chiral and anti-chiral superfields.

We exploit now the (anti-)BRST invariance that has been expressed in (C.7).
In fact, we have the following restrictions
\[\tilde E^{(h)}(x,\theta,\bar\theta)\cdot\bar B^{(r)}(x,\theta,\bar\theta) = E(x)\cdot\bar B(x),\quad \tilde E^{(h)}(x,\theta,\bar\theta)\cdot B^{(r)}(x,\theta,\bar\theta) = E(x)\cdot B(x),\eqno (C.10)\]
where the expansion for the $E^{(h)}(x,\theta,\bar\theta)$ is given in (C.4) and the reduced forms
of $\bar B^{(r)}(x,\theta,\bar\theta)$ and $B^{(r)}(x,\theta,\bar\theta)$ are quoted in (C.9).
Ultimately, with the substitution of these into (C.10), we obtain the following results, namely;
\[M(x) = i\;(B\times\bar C),\qquad\bar L(x) = i\;(\bar B\times C).\eqno (C.11)\]
Thus, we have the following explicit super expansions:
\[\tilde B^{(g)}(x,\theta,\bar\theta) = B(x) + \theta\; ( i \; B\times \bar C)\equiv B(x) + \theta\; (s_{ab} B),\]
\[\tilde {\bar B}^{(g)}(x,\theta,\bar\theta) = \bar B(x) + \bar\theta \;(i\; \bar B\times C)\equiv \bar B(x) + \bar\theta\; (s_b\bar B(x)).\eqno (C.12)\]
The above expressions for $\tilde B^{(g)}(x,\theta,\bar\theta)$ and $\tilde {\bar B}^{(g)}(x,\theta,\bar\theta)$
have been used in the super Lagrangian densities (C.1). Rest of the other terms in (C.1) are straightforward and clear.

We are now in the position to express the (anti-)BRST invariance of the Lagrangian densities (1)
 which change to the total spacetime derivatives under the above symmetry transformations (cf. Eqs. (3), (4)).
It is straightforward to check that 
\[\frac{\partial}{\partial\theta}\;\tilde {\cal L}_{\bar B}\Big|_{\bar\theta~ =~ 0} = -\;\partial_{\mu}(\bar B\cdot D^{\mu}\bar C),
\qquad \frac{\partial}{\partial\theta}\;\tilde {\cal L}_ B \Big|_{\theta~ = ~0}=  \;\partial_{\mu}( B\cdot D^{\mu} C),\eqno (C.13)\]
which are nothing but our earlier results (cf. Eq. (3)) where we have shown that 
$s_{ab}\;{\cal L}_{\bar B} = -\;(\bar B\cdot D^{\mu}\bar C)$   
and $s_b\; {\cal L}_B = \partial_{\mu} ( B\cdot D^{\mu}\bar C)$.
Geometrically, the above observations show that super Lagrangian densities (C.1)
are the sum of composite (super)fields,  obtained after (anti-)BRST invariant restrictions
and HC, such that their translation along the $(\theta,\bar\theta)$ directions of the 
(2, 2)-dimensional supermanifold produces the 
total spacetime derivatives.

In exactly similar fashion, we can discuss the (anti-)co-BRST invariance  of the 
Lagrangian densities (1) where these are generalized onto the (2, 2)-dimensional supermanifold as:
\[{\cal L}_B \longrightarrow \tilde {\cal L}_B = {\cal B}(x)\cdot \tilde E^{(dg)}(x,\theta,\bar\theta)
- \frac {1}{2}\;{\cal B}(x)\cdot  {\cal B}(x)   + B(x)\cdot\partial_{\mu}B^{\mu(dg)}(x,\theta,\bar\theta)\]
\[+ \frac {1}{2}\; (B(x)\cdot B(x) +  {\bar B}(x)\cdot {\bar B}(x))
- i\; \partial_{\mu}\bar F^{(dh)}(x, \theta, \bar\theta)\cdot \partial^{\mu} F^{(dh)}(x,\theta,\bar\theta)\]
\[+\; \partial_{\mu}\bar F^{(dh)}(x,\theta,\bar\theta)\cdot (B^{\mu(dg)}(x,\theta,\bar\theta)\times F^{(dh)}(x,\theta,\bar\theta)),\]

\[{\cal L}_{\bar B}\longrightarrow \tilde {\cal L}_{\bar B} = {\cal B}(x)\cdot \tilde E^{(dg)}(x,\theta,\bar\theta)
- \frac {1}{2}\;{\cal B}(x)\cdot  {\cal B}(x)   - \bar B(x)\cdot\partial_{\mu}B^{\mu(dg)}(x,\theta,\bar\theta)\]
\[+ \frac {1}{2}\; (B(x)\cdot B(x) +  {\bar B}(x)\cdot {\bar B}(x))
- i\; \partial_{\mu}\bar F^{(dh)}(x, \theta, \bar\theta)\cdot \partial^{\mu} F^{(dh)}(x,\theta,\bar\theta)\]
\[+\; (B_{\mu}^{(dg)}(x,\theta,\bar\theta)\times{\bar F}^{(dh)}(x,\theta,\bar\theta))\cdot\partial^{\mu}F^{(dh)}(x,\theta,\bar\theta),\eqno (C.14)\]
where the superscripts $(dh)$ and $(dg)$ on the superfields have already been explained in the main body of the text. We would like to comment 
here that the expression for $\tilde E^{(dg)}(x,\theta,\bar\theta)$ has been derived $(i. e.\,\,  F^{(dg)}_{01}(x,\theta,\bar\theta) = \tilde E^{(dg)}(x,\theta,\bar\theta))$ from the
superfield corresponding to the field strength tensor, namely;
\[\tilde F_{\mu\nu}^{(dg)}(x,\theta,\bar\theta) = \partial_\mu B_\nu^{(dg)}(x,\theta,\bar\theta) + \partial_\nu B_\mu^{(dg)}(x,\theta,\bar\theta) + i\,(B_\mu^{(dg)}(x,\theta,\bar\theta)\times
B_{\nu}^{(dg)}(x,\theta,\bar\theta)), \eqno (C.15)\]
where the expansion of $B_{\mu}^{(dg)}(x, \theta, \bar\theta)$ has been illustrated in Eq. (32).
In fact, the explicit substitution of this superfield into the above equation leads to the following:
\[\tilde E^{(dg)}(x, \theta, \bar\theta) = E(x) + \theta\; (D_{\mu}\partial^{\mu}C) + \bar\theta\; (D_{\mu}\partial^{\mu}\bar C) + \theta\;\bar\theta
\;(-\; i \; D_{\mu}\partial^{\mu}{\cal B} - i\;\varepsilon _{\mu\nu}\;(\partial^{\nu}\bar C\times \partial^{\mu}C))\]
\[\equiv  E(x) + \theta\; (s_{ad} E(x)) + \bar\theta\; (s_d E(x)) + \theta\;\bar\theta (s_d \;s_{ad}\; E(x)).\eqno (C.16)\] 
We note that the substitution of the super expansions from Eqs. (26) and (32) into the super Lagrangian densities (C.14) 
would express them in terms of the coefficients of $(1, \theta, \bar\theta, \theta\bar\theta)$.
It can be now checked that the following are true, namely;
\[\frac{\partial}{\partial\theta}\;\tilde {\cal L}_ {\bar B}\Big|_{\bar\theta~ =~ 0} = \partial_{\mu} [ {\cal B }\cdot\partial^{\mu} C]\quad\Longleftrightarrow \quad s_{ad}\; {\cal L}_ {\bar B} = \partial_{\mu}[{\cal B}\cdot\partial^{\mu} C],\]
\[\frac {\partial}{\partial\bar\theta}\;\tilde {\cal L}_ B\Big|_{\theta~ =~ 0} = \partial_{\mu} [ {\cal B }\cdot\partial^{\mu}\bar C]\quad\Longleftrightarrow \quad s_d\; {\cal L}_  B = \partial_{\mu}[{\cal B}\cdot\partial^{\mu}\bar C].\eqno (C.17)\]
Hence, we have provided the equivalence of the (anti-)co-BRST invariance of the Lagrangian densities (1) in the 
language of AVSA to BRST formalism. Consequently, the  (anti-)co-BRST  invariance can be explained within the framework 
of AVSA to BRST formalism as follows. The translation of the super Lagrangian densities (C.14) along $(\theta,\bar\theta)$-directions 
of the (2, 2)-dimensional supermanifold is such that it results in the total spacetime derivatives thereby 
rendering the action integrals (corresponding to the appropriate Lagrangian densities) invariant under 
the (anti-)co-BRST symmetry transformations. We end this {\it Appendix} with a concise remark that we can {\it also} capture the 
(anti-)co-BRST invariance of the coupled Lagrangian densities (39) exactly in the same manner as we have done for 
our starting Lagrangian densities (1) for the present 2D non-Abelian 1-form gauge theory.

\end{document}